\documentclass{article}
\usepackage[margin=1.0in]{geometry}

\usepackage[utf8]{inputenc} 
\usepackage[T1]{fontenc}    
\usepackage{hyperref}       
\usepackage{url}            
\usepackage{booktabs}       
\usepackage{amsfonts}       
\usepackage{nicefrac}       
\usepackage{microtype}      

\usepackage{authblk}

\usepackage{amsmath}
\usepackage{dsfont} 
\usepackage{graphicx}
\usepackage{caption}
\usepackage{subcaption}
\usepackage{float} 
\usepackage{tabularx}
\usepackage{longtable}
\usepackage{arydshln}
\usepackage{multirow} 
\usepackage[table]{xcolor}
\usepackage{pdflscape} 
\usepackage{flushend}
\usepackage{algorithm}
\usepackage{algorithmic}
\usepackage{tikz}
\usetikzlibrary{bayesnet} 

\usepackage[round,numbers,sort]{natbib}

\newcommand{\ie}{i.e., }

\newcommand{\eg}{e.g., }

\title{Characterizing physiological and symptomatic variation in menstrual cycles using self-tracked mobile-health data}

\author[$\dagger$,
		1,2]{Kathy Li (kathy.li@columbia.edu)}
\author[$\dagger$,
		1,2]{I\~nigo Urteaga (inigo.urteaga@columbia.edu)} 
\author[1,2]{Chris H.~Wiggins (chris.wiggins@columbia.edu)}
\author[3]{Anna Druet (anna@biowink.com)}
\author[3]{Amanda Shea (amanda.shea@biowink.com)}
\author[3,4]{Virginia J. Vitzthum (vitzthum@indiana.edu)}
\author[5,2]{No\'emie Elhadad (noemie.elhadad@columbia.edu)}

\affil[$\dagger$]{Equal contribution}

\affil[1]{Department of Applied Physics and Applied Mathematics, Columbia University, New York NY 10027}
\affil[2]{Data Science Institute, Columbia University, New York NY 10027}
\affil[3]{Clue by BioWink, Adalbertstra{\ss}e 7-8, 10999 Berlin, Germany}
\affil[4]{Kinsey Institute \& Department of Anthropology, Indiana University, Bloomington IN 47405}
\affil[5]{Department of Biomedical Informatics, Columbia University, New York NY 10032}

\begin{document}
\maketitle

\begin{abstract}
The menstrual cycle is a key indicator of overall health for women of reproductive age. Previously, menstruation was primarily studied through survey results; however, as menstrual tracking mobile apps become more widely adopted, they provide an increasingly large, content-rich source of menstrual health experiences and behaviors over time. By exploring a database of user-tracked observations from the Clue app by BioWink of over 378,000 users and 4.9 million natural cycles, we show that self-reported menstrual tracker data can reveal statistically significant relationships between per-person cycle length variability and self-reported qualitative symptoms. A concern for self-tracked data is that they reflect not only physiological behaviors, but also the engagement dynamics of app users. To mitigate such potential artifacts, we develop a procedure to exclude cycles lacking user engagement, thereby allowing us to better distinguish true menstrual patterns from tracking anomalies. We uncover that women located at different ends of the menstrual variability spectrum, based on the consistency of their cycle length statistics, exhibit statistically significant differences in their cycle characteristics and symptom tracking patterns. We also find that cycle and period length statistics are stationary over the app usage timeline across the variability spectrum. The symptoms that we identify as showing statistically significant association with timing data can be useful to clinicians and users for predicting cycle variability from symptoms or as potential health indicators for conditions like endometriosis. Our findings showcase the potential of longitudinal, high-resolution self-tracked data to improve understanding of menstruation and women's health as a whole.
\end{abstract}

\section{Introduction}
\label{sec:intro}
Menstruation is an important indicator of overall health and quality of life in women: the reproductive endocrine system is associated with sexual and reproductive health, bone and heart health, and cancers~\cite{popat_menstrual_2008, bedford_prospective_2010, zittermann_physiologic_2000, solomon_menstrual_2002,j-Carmina1999,j-Giudice2010,j-Barbosa2011, shuster_premature_2010, mahoney_shift_2010}; it affects fertility~\cite{j-Jordan1994, j-Crawford2017}, menopause~\cite{j-Prior1998,j-Landgren2004,j-Prior2011}, exercise~\cite{j-Prior1987}, and diet~\cite{j-Barr1995}. Seminal work on variation of menstrual cycle length throughout the reproductive lifespan~\cite{j-Arey1939,j-Treloar1967} has concluded that ``complete regularity in menstruation through extended time is a myth,'' and recent empirical studies~\cite{who1983, j-Ferrell2006,j-Gorrindo2007} have confirmed that variation between cycles, women, and populations is the norm~\cite{j-Chiazze1968,j-Muenster1992,j-Belsey1997,j-Burkhart1999,j-Vitzthum2000,j-Creinin2004,j-Williams2006,j-Vitzthum2009}. Establishing a clear, informative, and quantitative characterization of the patterns and underlying female physiology of what has been hypothesized as ``the fifth vital sign''~
\cite{acog2015vital,bobel_beyond_2019, lippe_scientific, american2006menstruation} has been a long-explored issue in women's health, but  remains an open research question~\cite{solomon2001long,solomon2002menstrual, j-Creinin2004}, in part due to limited access to large, reliable datasets concerning menstruation. 

With the rise of data-powered health, we now have the ability to identify menstrual patterns at scale and explore their relationships with a broad set of symptoms. Observational health data sources have shed light on individual clinical trajectories~\cite{j-Hripcsak2016}, increased self-awareness about individual health~\cite{li2010stage}, and helped deliver on the promise of precision medicine~\cite{j-Kohane2015}. Mobile-health solutions enable a high-resolution view of a large, highly diverse range of individuals over time~\cite{13pew_tracking,krebs2015health, j-Althoff2017a,j-Althoff2017} and can provide insights into chronic diseases and behaviors~\cite{chan2017asthma,j-Webster2018, j-Egger2018, bot2016mpower, j-Dagum2018,j-Smets2018,j-Byambasuren2018, j-Ata2018,j-Torous2018,ip-Urteaga2018}. Menstrual trackers in particular have become increasingly common: they are the second most popular app for adolescent girls and the fourth most popular for adult women~\cite{j-Wartella2016,Fox2012}. Millions of women around the world routinely track their menstrual cycles and a variety of contextual factors and symptoms, accumulating high volumes of temporal, heterogeneous data via many different apps~\cite{clue_app,dot_app,glow_app,spoton_app,natural_cycle_app}. As exemplified by studies connecting the menstrual cycle to variations in women's mood, behavior, and vital signs~\cite{j-Pierson2019}, self-tracked data can provide insights into cycle characteristics~\cite{j-Bull2019}, ovulation timing, and the evolution of reproductive health for large populations~\cite{j-Symul2019}, as well as empower informed decision-making through increased self-awareness~\cite{j-Epstein2017}.

\begin{figure}[h]
	\centering
	\includegraphics[width=0.9\linewidth]{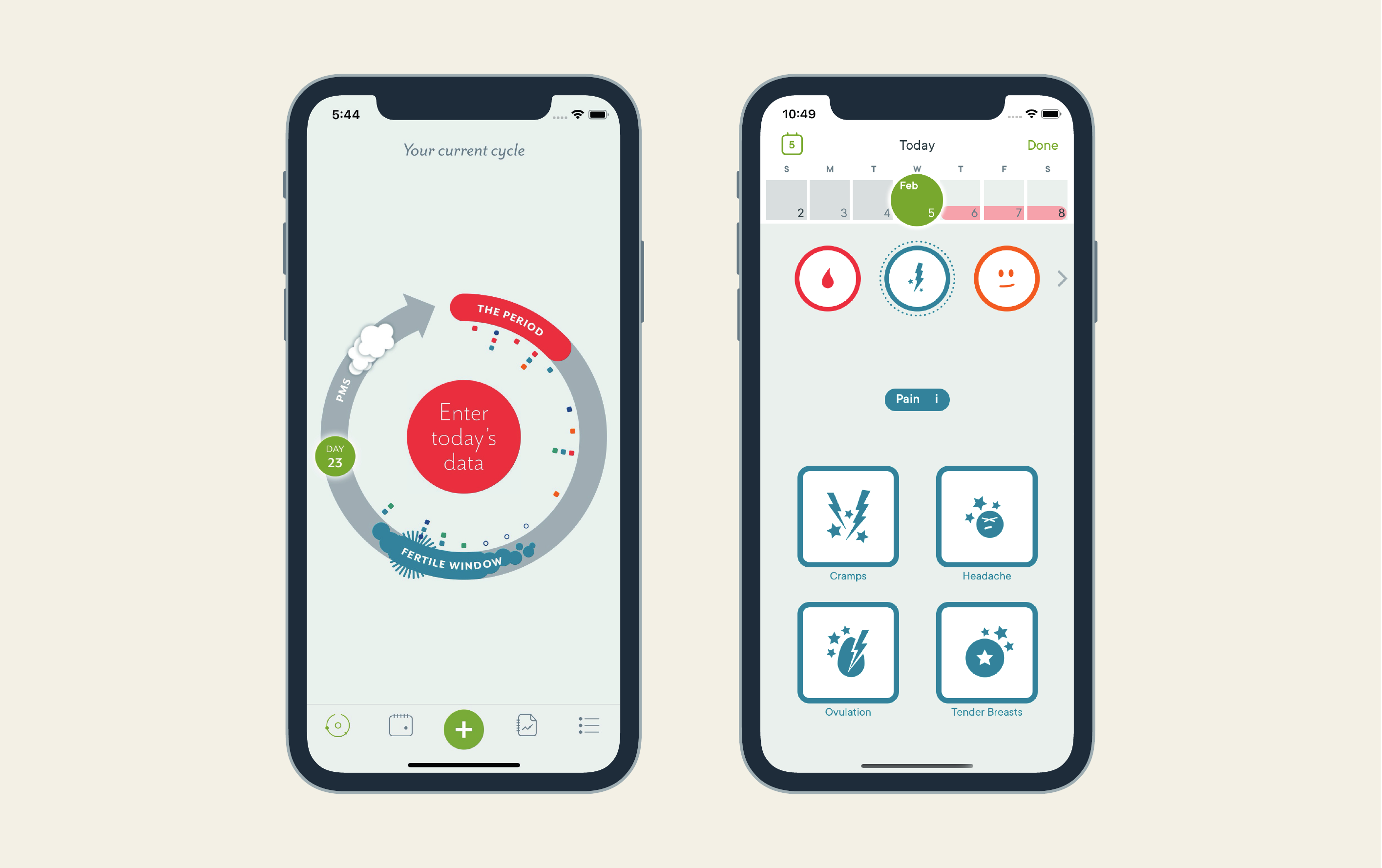}
	\caption{Sample screenshots of the Clue app. Users can track daily symptoms across 20 categories; Table~\ref{tab:statistics_symptoms} provides a description of the available Clue categories and their corresponding symptoms. On the left for example, the app displays what day the user is currently on in their cycle. On the right, a user can choose from `cramps,' `headache,' `ovulation,' or `tender breasts' symptoms for the category `pain' (the third most tracked category in our dataset, see Table~\ref{tab:statistics_symptoms}).}
	\label{fig:clue_screenshot}
	\vspace*{-2ex}
\end{figure}

We utilize de-identified user-tracked data from Clue by BioWink~\cite{clue_app}, one of the most popular and accurate menstrual trackers worldwide~\cite{moglia_evaluation_2016}. In addition to period data, Clue users can track symptom information in categories like exercise, pain, and sexual activity (see Figure~\ref{fig:clue_screenshot}). Note that Clue users are not required to specify gender---in this paper, we refer to Clue users or menstruators as `women,' but we acknowledge that not all menstruators are women and vice versa. This large-scale dataset provides a high resolution, long-term view of variation in both physiology (period and cycle duration) and symptoms (\eg pain and mood) across menstrual cycles, enabling us to study the shared information between quantitative, temporal attributes and qualitative, symptomatic attributes of menstrual experiences. 

While previous work has examined how menstrual cycle characteristics like cycle and phase length vary with age and body mass index (BMI)~\cite{j-Bull2019}, we aim instead to use the observed variability in cycle length statistics to investigate differences in symptomatic behavior between those who exhibit more or less variable cycle lengths. Namely, we seek to answer two research questions:

\textbf{(1)} how do cycle length characteristics for a large, self-tracked user population differ among groups of users?;
and \textbf{(2)} how do users who fall at different ends of the cycle length variability spectrum self-track their menstrual symptoms?

To this end, we select users from the Clue dataset aged 21-33 (because menstrual cycle lengths are relatively less variable and cycles are more likely to be ovulatory during this age interval~\cite{j-Treloar1967,j-Chiazze1968,j-Ferrell2005,j-Vitzthum2009,j-Harlow2012}) with natural menstrual cycles (\ie no hormonal birth control or intrauterine device (IUD)). We define a menstrual \textbf{cycle} as the span of days from the first day of a period through to and including the day before the first day of the next period~\cite{j-Vitzthum2009}. A \textbf{period} consists of sequential days of bleeding (greater than spotting and within ten days after the first greater-than-spotting bleeding event) unbroken by no more than one day on which only spotting or no bleeding occurred. 

In this paper, we take symptom tracking behavior to be a proxy for true physiological behavior. The Clue tracking categories (summarized in Table~\ref{tab:statistics_symptoms}) encompass a wide range of experiences (subject to user interpretation of the category rather than based on specific validated scales), enabling broad usage of the app to meet individual user needs. Self-tracked data reliability is dependent on consistent and accurate user tracking; for instance, cycle length can be arbitrarily long if a user forgets to track their period, which would skew the analysis of menstrual patterns by misrepresenting a long cycle as due to physiological behavior rather than tracking behavior. We propose a procedure to mitigate such potential engagement artifacts by quantifying engagement with cycle tracking and identifying cycles lacking engagement, allowing us to separate true menstrual patterns from tracking anomalies. To investigate the spectrum of variability in women's menstrual health experiences, we propose \textbf{cycle length difference} or CLD---the absolute difference between subsequent cycle lengths---as a robust metric for quantifying cycle variability, and we examine users who fall at opposite ends of the variability spectrum.
\section{Results}
\label{sec:results}

\paragraph*{Study population.} The cohort for this study comprises $378,694$ users located on all continents, aged 21 to 33 years old (see Table~\ref{tab:summary_statistics} for detailed summary statistics). The average user is $25.49$ (median of $25$) years old (per-country and per-age detailed statistics are provided in 
\ifx\includesi\undefined \href{https://github.com/iurteaga/menstrual_cycle_analysis/blob/master/doc/characterization/supplementary_information.pdf}{the Supplementary Information}\else the Supplementary Information\fi
). As reported in Table~\ref{tab:statistics_cycles}, the average number of cycles tracked per user is $12.89$ (median of $11$), with an average cycle length of $29.73$ (median of $29$) days and mean period length of $4.08$ (median of $4$) days.

\begin{table*}[!h]
\caption{Summary statistics of the full cohort}
\label{tab:summary_statistics}
\begin{center}
\begin{tabularx}{\textwidth}{|X|c|c|c|}
	\hline
	Variable & Full cohort & Consistently not highly variable & Consistently highly variable \\ \hline
Number of users & 378,694 (100.00\%) & 349,606 (92.32\%) & 29,088 (7.68\%)\\
\hdashline
Number of observations & 117,014,597 (100.00\%) & 112,093,683 (95.79\%) & 4,920,914 (4.21\%)\\
\hdashline
Number of days of observation & 34,056,343 (100.00\%) & 32,699,312 (96.02\%) & 1,357,031 (3.98\%)\\
\hdashline
Number of cycles & 4,881,697 (100.00\%) & 4,701,694 (96.31\%) & 180,003 (3.69\%)\\
\hline
\end{tabularx}
\end{center}
Summary statistics of the full cohort, as well as for the consistently not highly variable and consistently highly variable user groups. We utilize a greater than 9 day median cycle length difference threshold to place users in each group---those in the consistently highly variable group represent the far end of a cycle variability spectrum.
\end{table*}

\begin{table*}	
\caption{Description of the Clue app tracking categories and symptoms}
\label{tab:statistics_symptoms}
\begin{center}
	\begin{tabularx}{1\textwidth}{|X|X|X|X|X|}
		\hline
		Category & Description & Symptoms & Number of tracking events (\%) for the consistently not highly variable group & Number of tracking events (\%) for the consistently highly variable group \\ \hline
		period  &  Period flow  &  spotting, light, medium, heavy  & 22,096,884 (19.71\%) & 913,403 (18.56\%) \\ \hdashline
		emotion  &  Emotional state  &  happy, sensitive, sad, PMS  & 11,377,997 (10.15\%) & 501,610 (10.19\%) \\ \hdashline
		pain  &  Type of pain experienced  &  cramps, tender breasts, headache, ovulation pain  & 9,730,958 (8.68\%) & 406,710 (8.26\%) \\ \hdashline
		energy  &  Energy level  &  low, high, exhausted, energized  & 8,710,403 (7.77\%) & 410,216 (8.34\%) \\ \hdashline
		sleep  &  Hours of sleep  &  0-3, 3-6, 6-9, > 9  & 8,597,769 (7.67\%) & 405,726 (8.24\%) \\ \hdashline
		skin  &  Skin health  &  acne, good, oily, dry  & 5,896,540 (5.26\%) & 263,258 (5.35\%) \\ \hdashline
		mental  &  Mental state  &  calm, distracted, focused, stressed  & 5,871,137 (5.24\%) & 252,621 (5.13\%) \\ \hdashline
		sex  &  Sexual health  &  unprotected sex, high sex drive, protected sex, withdrawal sex  & 5,813,292 (5.19\%) & 271,540 (5.52\%) \\ \hdashline
		motivation  &  Motivation level  &  motivated, unmotivated, productive, unproductive  & 5,467,728 (4.88\%) & 236,052 (4.80\%) \\ \hdashline
		craving  &  Food cravings  &  sweet, salty, carbs, chocolate  & 4,867,777 (4.34\%) & 224,751 (4.57\%) \\ \hdashline
		digestion  &  Digestive health  &  great, bloated, gassy, nauseated  & 4,825,627 (4.30\%) & 209,651 (4.26\%) \\ \hdashline
		social  &  Social behavior  &  sociable, withdrawn, supportive, conflict  & 4,178,744 (3.73\%) & 186,110 (3.78\%) \\ \hdashline
		poop  &  Stool health  &  normal, constipated, great, diarrhea  & 3,889,471 (3.47\%) & 172,716 (3.51\%) \\ \hdashline
		hair  &  Hair health  &  good, bad, oily, dry  & 3,128,384 (2.79\%) & 147,844 (3.00\%) \\ \hdashline
		fluid  &  Vaginal discharge type  &  creamy, egg white, sticky, atypical  & 2,378,211 (2.12\%) & 106,782 (2.17\%) \\ \hdashline
		collection method  &  Method for period collection  &  pad, tampon, panty liner, menstrual cup  & 2,027,258 (1.81\%) & 84,270 (1.71\%) \\ \hdashline
		exercise  &  Physical exercise  &  running, yoga, biking, swimming  & 1,222,568 (1.09\%) & 44,946 (0.91\%) \\ \hdashline
		party  &  Party-related experiences  &  drinks, cigarettes, big night, hangover  & 900,444 (0.8\%) & 40,779 (0.83\%) \\ \hdashline
		medication  &  Type of medication taken  &  pain, cold / flu, antihistamine, antibiotic  & 561,540 (0.5\%) & 21,030 (0.43\%) \\ \hdashline
		ailment  &  Physical maladies  &  cold / flu, allergy, injury, fever  & 550,951 (0.49\%) & 20,899 (0.42\%) \\ \hline
	\end{tabularx}
\end{center}
Description of tracking categories and corresponding symptoms for the Clue app, along with the per-symptom number of tracking observations (and their corresponding proportion with respect to the total number of observations) for the consistently not highly variable and consistently highly variable user groups.
\end{table*}

\begin{landscape}
\begin{table*}
\caption{Per-user cycle characteristics}
\label{tab:statistics_cycles}
\begin{center}
\begin{tabular}{|c|c|c|c|}
	\hline
	 & Full cohort's & Consistently not highly variable group's & Consistently highly variable group's \\
	Variable & mean$\pm$sd, (95\% CI), median & mean$\pm$sd, (95\% CI), median & mean$\pm$sd, (95\% CI), median \\
	\hline
Number of cycles & 12.89 $\pm$ 9.11 (3.00,36.00) 11.00 & 13.45 $\pm$ 9.19 (3.00,37.00) 11.00 & 6.19 $\pm$ 3.87 (2.00,17.00) 5.00 \\
\hdashline
Cycle length & 29.73 $\pm$ 5.73 (21.00,43.00) 29.00 & 29.45 $\pm$ 4.98 (21.00,41.00) 29.00 & 37.04 $\pm$ 13.71 (13.00,69.00) 34.00 \\
\hdashline
Period length & 4.08 $\pm$ 1.76 (1.00,7.00) 4.00 & 4.07 $\pm$ 1.72 (1.00,7.00) 4.00 & 4.28 $\pm$ 2.54 (1.00,9.00) 4.00 \\
\hdashline
Median CLD & 4.15 $\pm$ 4.94 (1.00,18.00) 3.00 & 3.04 $\pm$ 1.86 (1.00,8.00) 2.50 & 17.48 $\pm$ 9.15 (9.50,43.00) 14.00 \\
\hdashline
Maximum CLD & 10.07 $\pm$ 7.49 (2.00,31.00) 8.00 & 8.82 $\pm$ 5.65 (2.00,23.00) 8.00 & 25.15 $\pm$ 10.10 (12.00,53.00) 23.00 \\
\hline
\end{tabular}
\end{center}
Per-user high-level cycle characteristics for the full cohort, as well as for the consistently not highly variable and consistently highly variable user groups. We utilize a greater than 9 day median cycle length difference threshold to place users in each group---those in the consistently highly variable group represent the far end of a cycle variability spectrum. The `cycle length difference' (CLD) refers to the absolute difference between two consecutive cycles.
\end{table*}

\end{landscape}

\paragraph*{Cycle length difference (CLD) as a robust metric for quantifying cycle variability.}
We propose cycle length difference, or CLD---the absolute difference between subsequent cycle lengths---as a powerful metric to characterize the spectrum of menstrual variability.
We examine each user's CLDs to identify those who are `\textit{consistently highly variable}' in their cycle lengths. We find that a median CLD of 9 days splits consistently highly variable and consistently not highly variable cycle behavior, and we use this threshold to separate the menstrual experiences and symptom reporting of those at different ends of the variability spectrum. Tables ~\ref{tab:summary_statistics}, ~\ref{tab:statistics_cycles}, and ~\ref{tab:statistics_symptoms} showcase the summary statistics, high-level cycle characteristics, and category tracking frequencies for the resulting two groups, respectively. We note that the consistently highly variable group comprises about 7.68\% of the user cohort (29,088 out of 378,694 users) and that their relative category tracking frequencies is similar to the larger, consistently not highly variable user group. Period flow, emotional state, and experienced pain are the most frequently tracked categories across both groups; they account for 38.54\% of the events for the consistently not highly variable group and 37.01\% for the consistently highly variable group. Below, we summarize the commonalities and differences in \textbf{cycle and period length characteristics} and \textbf{symptom tracking behavior} between these two populations.


\begin{figure}[!h]
	\centering
	\includegraphics[width=0.5\columnwidth]{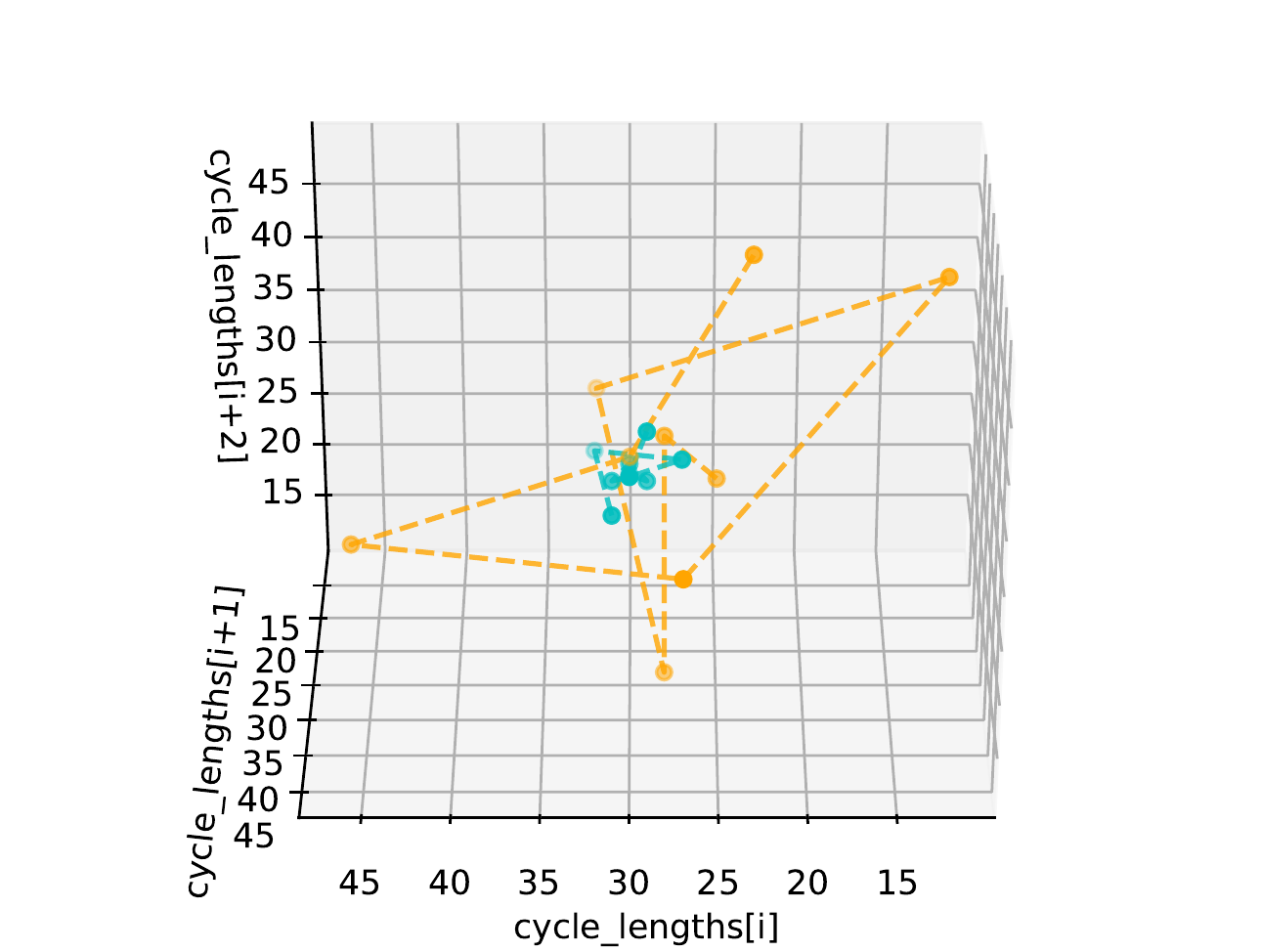}
	\caption{We sample one consistently highly variable and one consistently not highly variable user, each with the median number of cycles (11), from the user cohort and plot each set of three consecutive cycles on the $x, y$ and $z$ axes, respectively. This allows us to visualize how much a user's cycle lengths change throughout their entire cycle tracking history---we would expect that a not consistently highly variable user would have points that cluster closer together in space. We see that the consistently not highly variable (teal) user occupies a small region, while the consistently highly variable (orange) user’s points move through the space. This indicates that the teal user's cycle lengths are consistently very similar to one another, whereas the orange user experiences more consistent fluctuation in cycle lengths. Thus, we see that separating users into groups on the basis of median CLD identifies those who are more and less consistently highly variable.}
	\label{fig:time_series_sample}
	\vspace*{-2ex}
\end{figure}

\begin{figure*}[!h]
	\centering
	\includegraphics[width=0.45\textwidth]{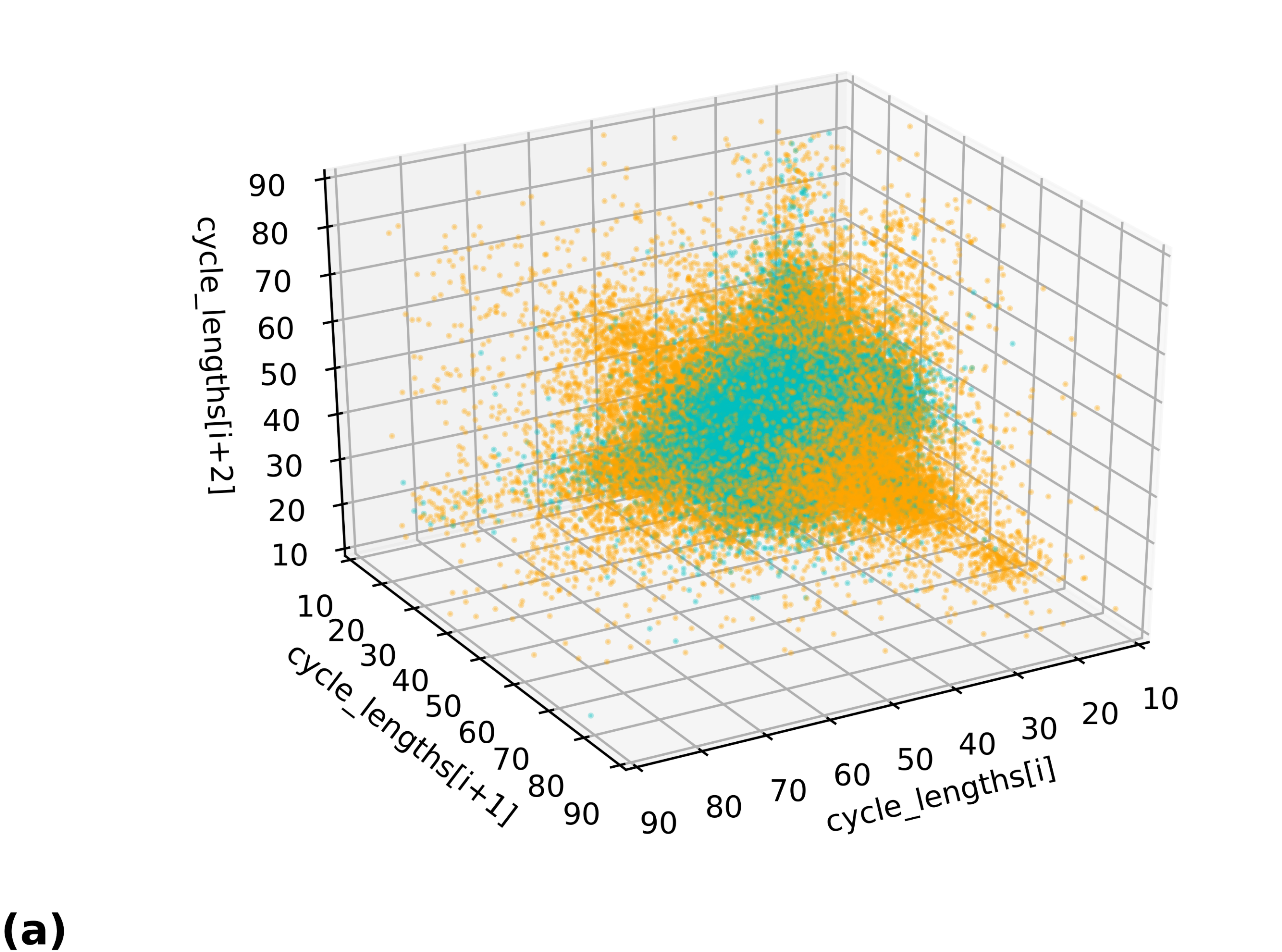}
	\includegraphics[width=0.45\textwidth]{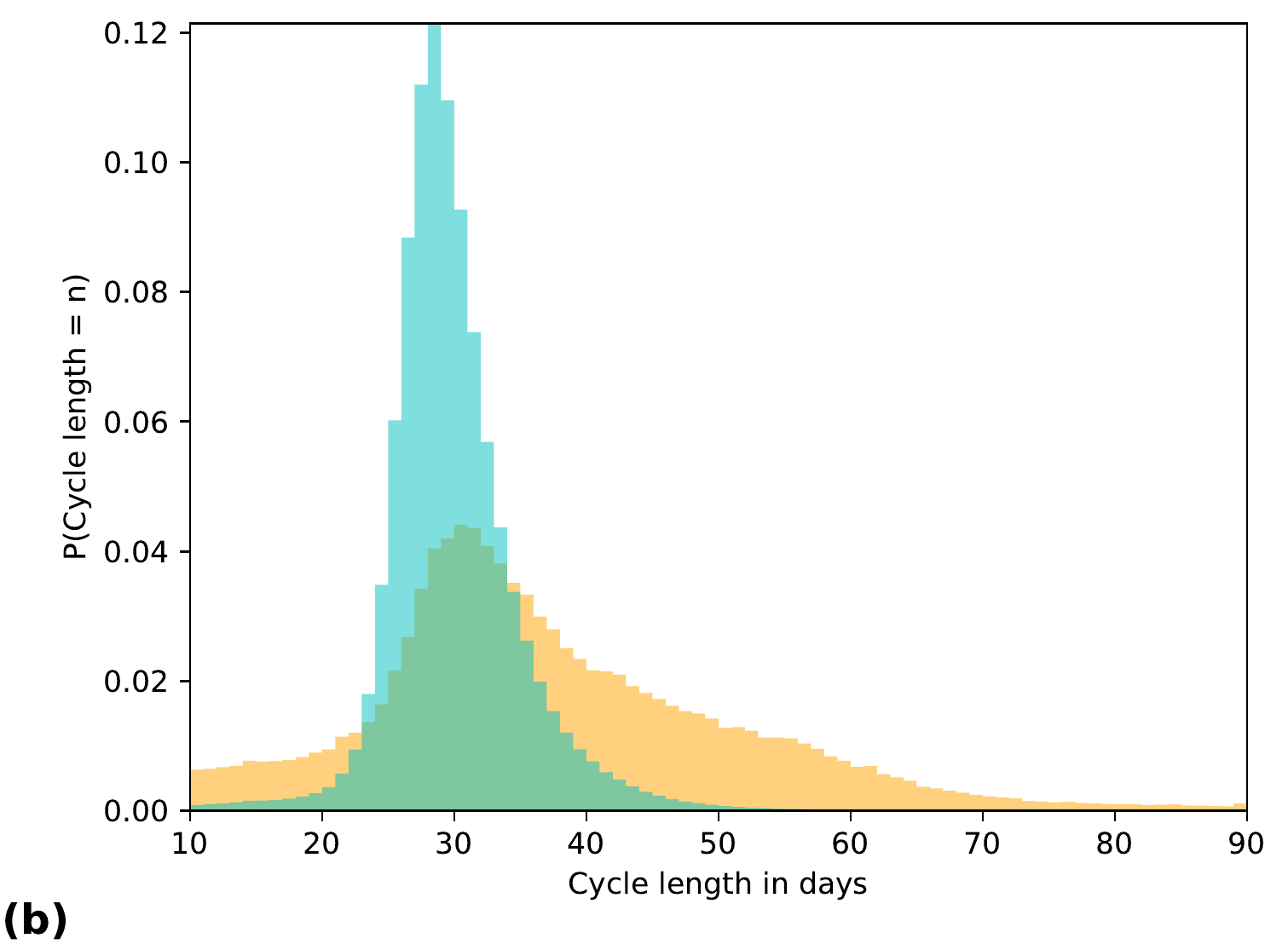}
	\caption{
		Time series embedding (\textbf{a}) and probability distributions (\textbf{b}) of cycle length for the consistently not highly variable (teal) and consistently highly variable (orange) groups.
		\textbf{(a)} The cycle lengths of three consecutive randomly sampled cycles from each user in the cohort are plotted on the $x, y,$ and $z$ axes. Each consistently not highly variable user is represented by a teal point, and each consistently highly variable user by an orange point. It is visually evident that the teal cluster of users occupies a tighter region of the space around the $x=y=z$ line, with the orange cluster fanning outward. 
		\textbf{(b)} The cycle length probability distributions of the cohort, where we note that the orange group's distribution has a much wider spread and is less peaked than the teal group. Cycle lengths are more heterogeneous or widely distributed for the orange group, confirming that the consistently highly variable group represents those with more fluctuation in cycle length. The cumulative distributions per-group differ significantly (as per a two-sample KS test).
	}
	\label{fig:cycle_length}
	\vspace*{-2ex}
\end{figure*}

\paragraph*{Cycle variability characterization --- Women in the consistently highly variable group experience volatile cycle lengths.}
We examine the cycle length characteristics of the proposed user groups, both visually and statistically. At an individual level, we visualize for two randomly sampled users (one per group) a time series embedding of all of their consecutive cycle lengths in Figure~\ref{fig:time_series_sample}. We sample one consistently highly variable and one consistently not highly variable user with the median number of cycles (11) from the cohort and plot each set of three consecutive cycles on the $x, y$ and $z$ axes, respectively. In Figure~\ref{fig:time_series_sample}, the consistently not highly variable (teal) user occupies a small region of the space, indicating that this user experiences similar cycle lengths throughout their history; however, the consistently highly variable (orange) user's points wander through the space, indicating that this user experiences consistently fluctuating cycle lengths throughout their cycle history.

In Figure~\ref{fig:cycle_length}a, we plot the time series embeddings of cycle length for the entire cohort, where each point represents three consecutive cycles randomly sampled from each user's cycle history (for users with at least three cycles). In contrast to Figure \ref{fig:time_series_sample}, each user is represented by one point, instead of plotting the whole cycle histories of two randomly sampled users. We visualize at a population level whether our median CLD metric successfully separates out groups of users based on their cycle length fluctuations. If a user is perfectly consistently not highly variable, then its representative point would fall exactly on the $x = y = z$ line, since the three cycle lengths would be identical (i.e., not fluctuating at all). We observe a consistent phenomenon in Figure~\ref{fig:cycle_length}a: the consistently not highly variable group (teal) occupies a tighter region of the space than the consistently highly variable one (orange). That is, a user in the consistently highly variable group experiences volatile menstrual patterns (i.e., highly varying cycle lengths).

Furthermore, we study the empirical cycle length distributions per group, and as seen in Figure~\ref{fig:cycle_length}b, the cycle length distributions differ significantly between the two user groups. Observe that not only are cycle length statistics such as mean and median cycle length different, but that the shapes of the distributions are also distinct. Specifically, in addition to being centered at longer cycle lengths (median of 34 days versus 29 days), the cycle length distribution for the consistently highly variable group is less peaked with a wider spread (encompasses a more volatile range of cycle lengths), has much heavier tails, and is skewed towards longer cycle lengths.

\begin{figure*}[!h]
	\centering
	\includegraphics[width=0.45\textwidth]{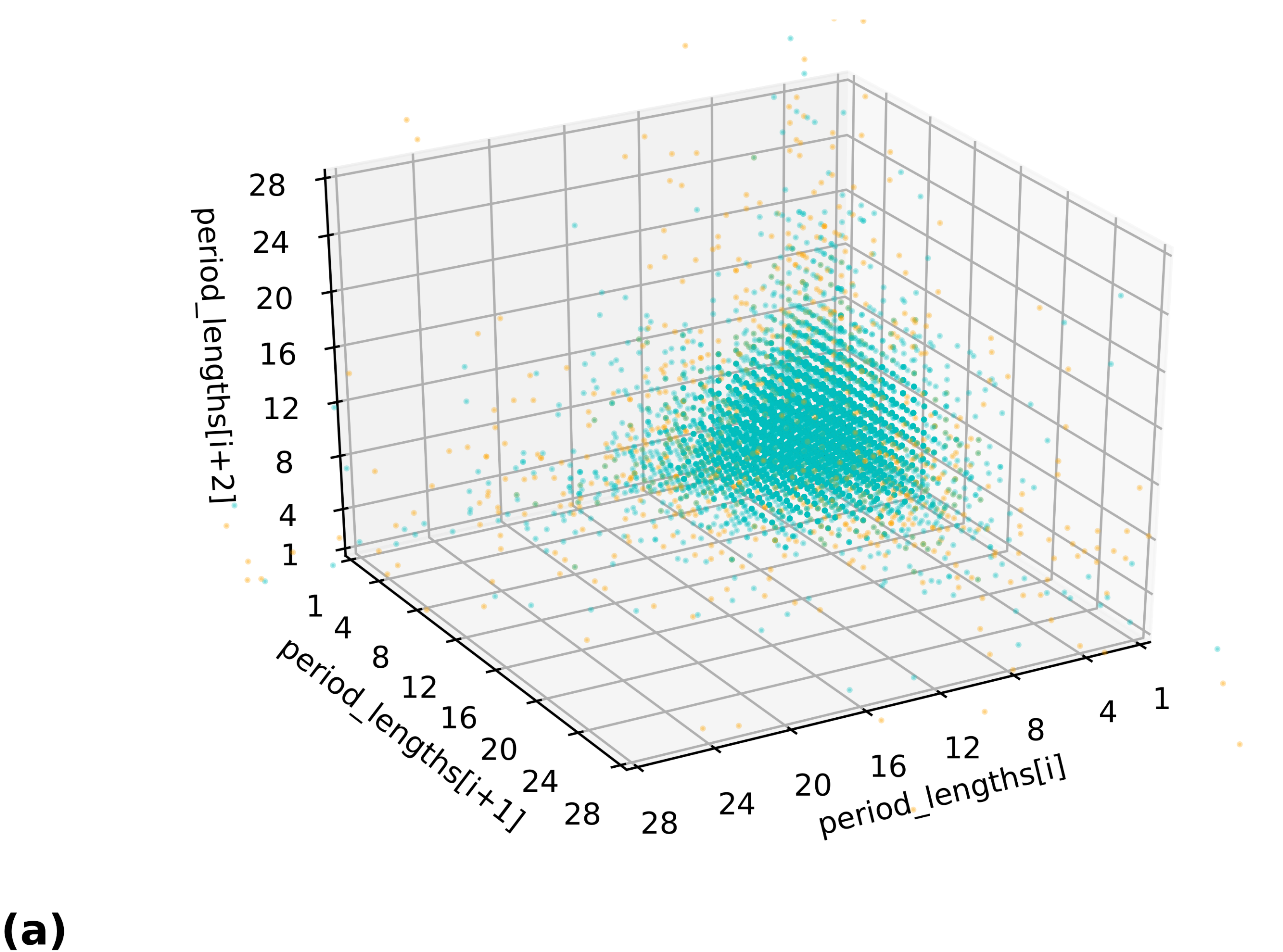}
	\includegraphics[width=0.45\textwidth]{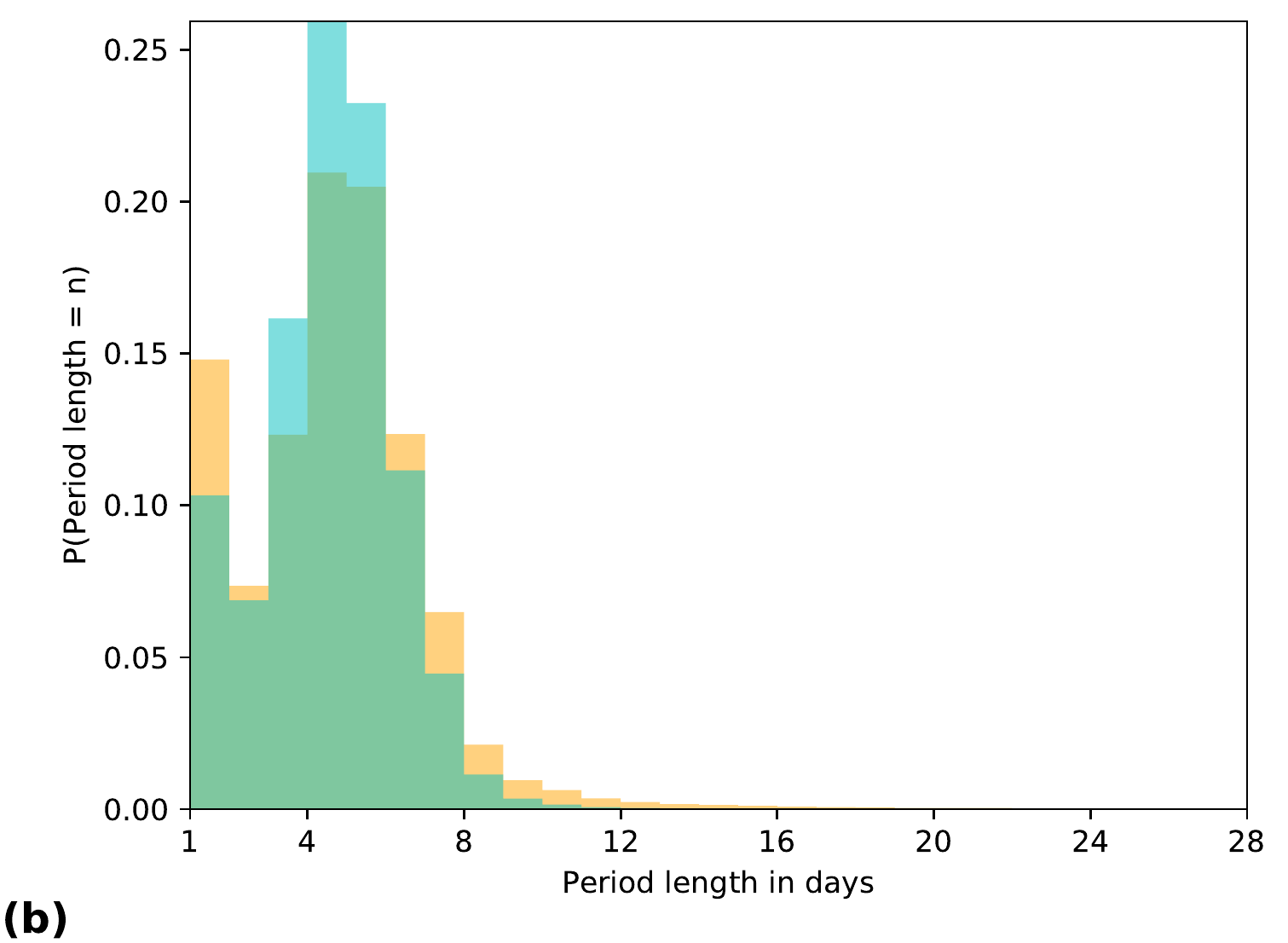}
	\caption{Time series embedding (\textbf{a}) and probability distributions (\textbf{b}) of period length for the consistently not highly variable (teal) and consistently highly variable (orange) groups.
		\textbf{(a)} The period lengths of three consecutive randomly sampled cycles from each user in the cohort are plotted on the $x, y,$ and $z$ axes. Visually, we observe that both groups occupy a very similar region of the period length space (few orange points are placed outside the region occupied by the teal cluster). 
		\textbf{(b)} The period length probability distributions of the cohort, where we observe that the orange and teal distributions are largely overlapping, with the same median of 4 days and a similar shape, indicating that period lengths are distributed very similarly for the two groups.
		We notice a slight peak in single day period reports in both groups, which we argue is reminiscent of app usage behavior: some users are interested in knowing (approximately) when they had their period, not in tracking how long it was, so they may only track the day it occurred and not continue tracking after that.
	}
	\label{fig:period_length}
\end{figure*}

\paragraph*{Cycle variability characterization---Period length statistics are homogeneous across the variability spectrum.}
We find that while women in different groups as separated by median CLD differ in their cycle length variability, their period length distributions are much less variable and fluctuate similarly between the groups. Period length is centered around the same median of 4 days for both groups and displays a similar length distribution. Figure~\ref{fig:period_length} confirms that the variability in cycle length is not due to period length differences between the groups, as the period length varies the same amount across all women. These results show that our metric (median CLD) identifies two distinct groups of users based on their cycle (not period) length variability. Note that while the period length distributions do differ significantly by the two-sample Kolmogorov-Smirnov (KS) test, the KS statistic for the period length distributions is 0.066 with a 95\% confidence interval of (0.064, 0.068) (details on computing the confidence interval using bootstrapping are presented in the Methods section). These numbers are nearly an order of magnitude smaller than those for the cycle length distributions (0.377 with a 95\% confidence interval of (0.375, 0.378)). That is, the KS test identifies the cycle length distributions to differ more drastically and with much higher probability than the period length distributions.

\begin{figure*}[!h]
	\centering
	\includegraphics[width=0.45\textwidth]{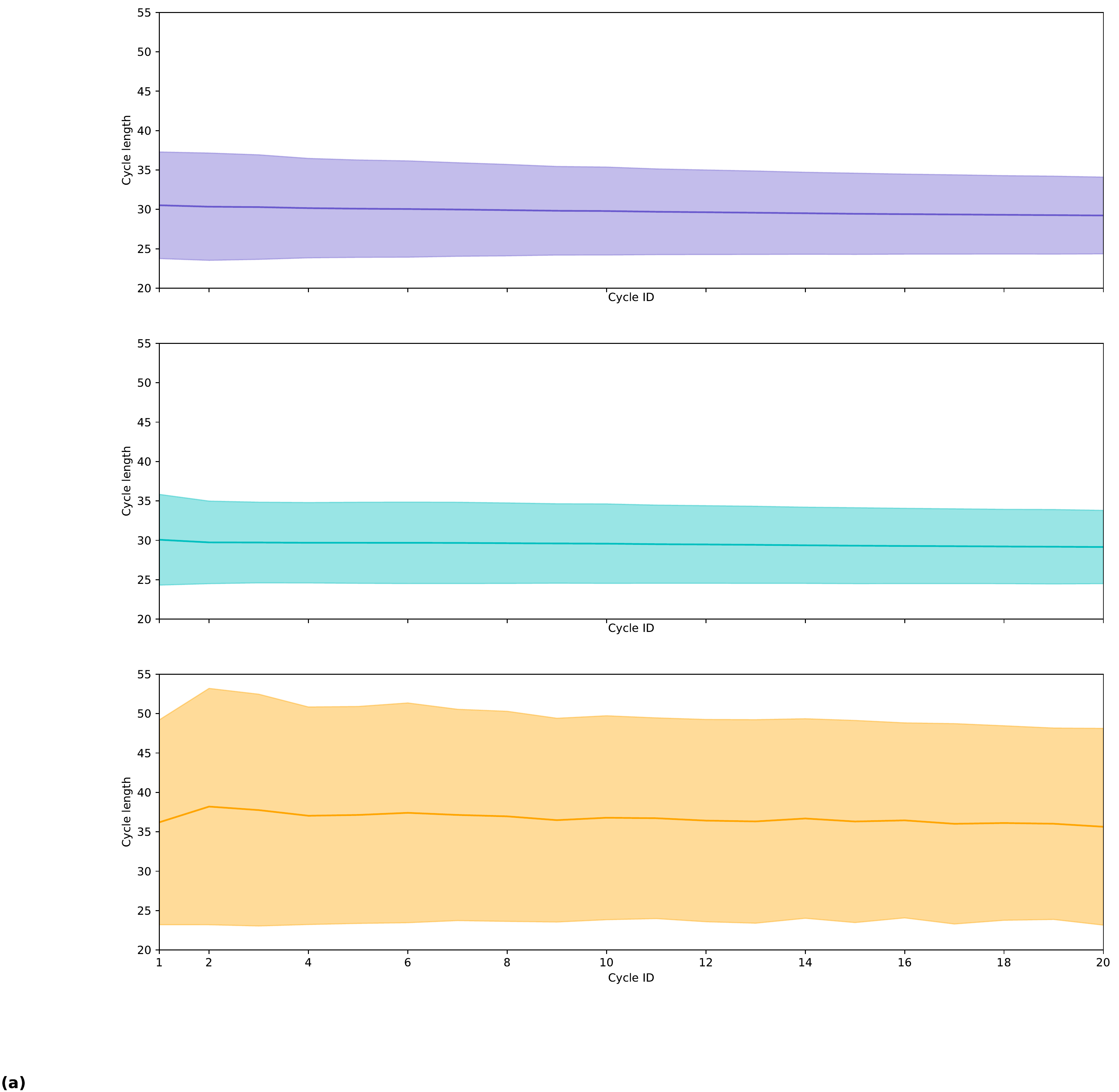}
	\includegraphics[width=0.45\textwidth]{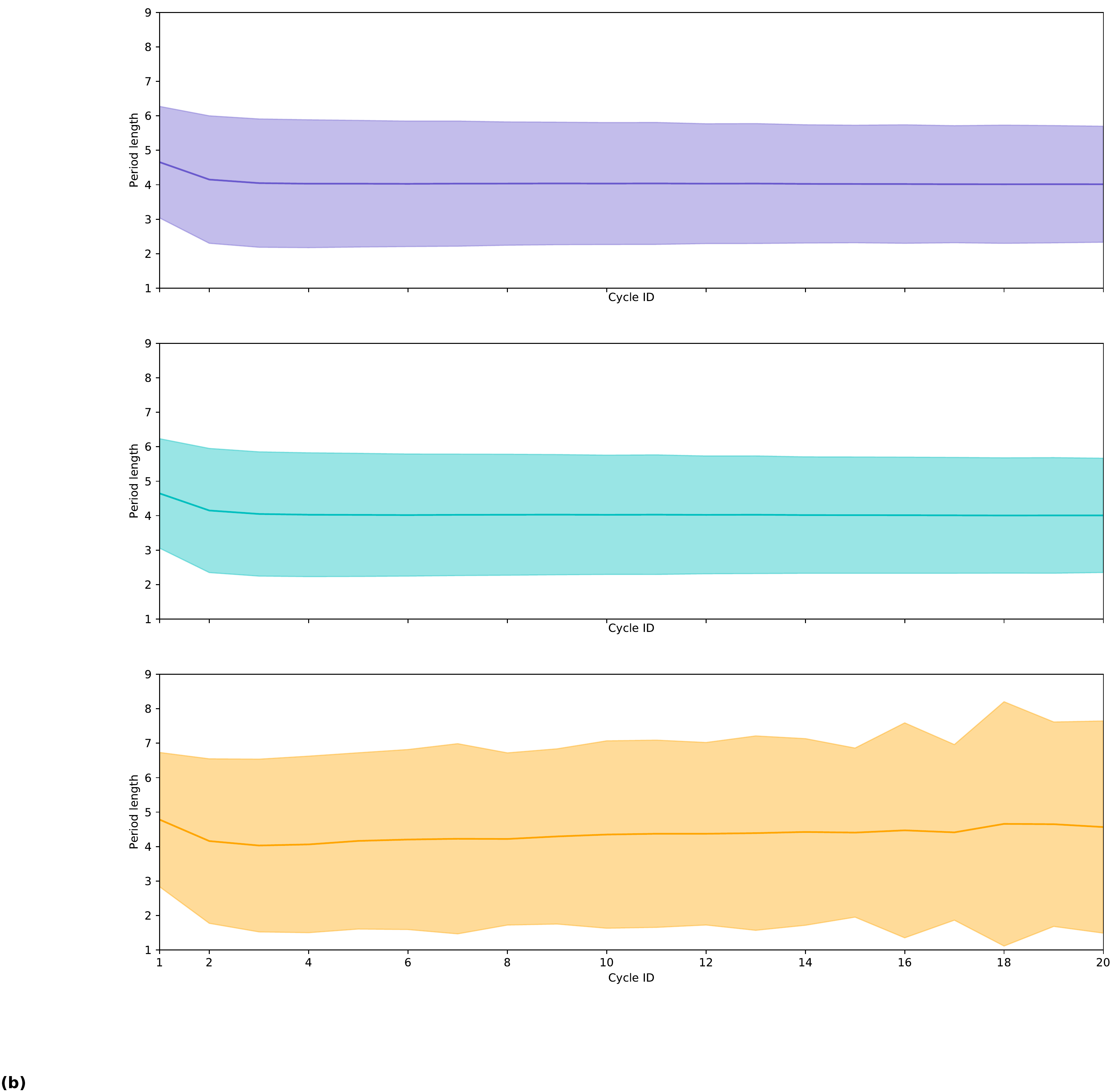}
	\caption{
			For each user's cycles (indexed by cycle ID), we average cycle (\textbf{a}) and period length (\textbf{b}) across three different groups: the entire user cohort (top, purple), the consistently not highly variable user cohort (middle, teal), and the consistently highly variable user cohort (bottom, orange). This allows us to visualize how cycle and period length vary over time for each group on average and in terms of standard deviation (for illustrative purposes, we restrict the cycle ID to 20). Cycle and period length statistics are stationary over the app usage timeline within each plot. We note that the top and middle plots look similar in each figure (i.e., the consistently not highly variable group looks similar to the overall population in terms of both cycle and period length), but the wider shaded orange spread of the bottom plot demonstrates the higher degree of variability in the consistently highly variable group. In addition, this spread is consistently wider for the orange plot over time. This showcases that the consistently highly variable group represents a large degree of the variability that we see in the data overall.}
	\label{fig:avg_cycle_period_length_over_time}
\end{figure*}

\paragraph*{Cycle variability characterization---Cycle and period length statistics are stationary within groups over the app usage timeline.}
We study per-group cycle statistics over the app usage time (as represented by cycle ID) in Figure~\ref{fig:avg_cycle_period_length_over_time} and find that cycle and period length statistics are stationary over time at the group level. Cycle ID enables us to align all users according to their subsequently tracked cycles (not absolute time), \ie a cycle ID of 1 corresponds to the first cycle of a user, 2 to their second cycle, and so on. As reported in Table~\ref{tab:statistics_cycles}, the mean cycle length for the consistently not highly variable group is 29.45 days (median of 29), and the mean is 37.04 days (median of 34) for the consistently highly variable group.
We observe that while average cycle and period length are similar over subsequent reported cycles for both the entire user cohort and the consistently not highly variable user group, consistently highly variable users exhibit a wider spread (\ie higher volatility). This volatility is maintained across cycles for users in the consistently highly variable group, showcasing that this group accounts for a large degree of the volatility in the data; this detail would be largely `smoothed out' and lost if we considered the whole population rather than separating the users into two groups.
Since cycle and period length statistics are constant within groups across app usage, we are confident that the proposed median CLD is not merely capturing spurious correlations that depend on how long the user stays with the app.

\paragraph*{Reported symptom differences---Women located at different ends of the spectrum of menstrual variability exhibit different symptom patterns.}
We find that there exists a relationship between median CLD and cycle level user symptom tracking behavior---despite CLD being a measure of cycle length variability, it is also correlated with symptom tracking behavior. Specifically, our analysis of the symptoms tracked across the two variability groups showcases that while users exhibit similar tracking frequencies (\ie the total number of times they track throughout history) per category (as in Table~\ref{tab:statistics_symptoms}), there are notable differences among their symptom tracking patterns (\ie how they track throughout history). The population level distributions of our metric (\ie `proportion of cycles with symptom out of cycles with category' in Equation~\ref{eq:our_metric}) differ between the user groups across most categories, with these differences being significant for all symptoms within the period, pain, and emotion categories, a result which may be clinically useful for assessing menstrual conditions and overall wellness. We present the KS test results for symptoms within those categories in Table~\ref{tab:interesting_symptoms_ks}.

\begin{table*}
\caption{Kolmogorov-Smirnov test results for symptom tracking patterns}
\label{tab:interesting_symptoms_ks}
\begin{center}
\begin{tabular}{|c|c|c|}
\hline
Category & Symptom & Kolmogorov-Smirnov statistic (95\% CI) \\
\hline
Period flow&heavy&0.181 (0.178,0.183) \\ \hdashline
Period flow&medium&0.134 (0.132,0.137) \\ \hdashline
Period flow&light&0.121 (0.118,0.124) \\ \hdashline
Period flow&spotting&0.089 (0.087,0.092) \\ \hdashline
Type of pain experienced&cramps&0.101 (0.097,0.104) \\ \hdashline
Type of pain experienced&ovulation pain&0.096 (0.093,0.099) \\ \hdashline
Type of pain experienced&headache&0.089 (0.087,0.092) \\ \hdashline
Type of pain experienced&tender breasts&0.082 (0.080,0.084) \\ \hdashline
Emotional state&sensitive emotion&0.115  (0.112,0.118) \\ \hdashline
Emotional state&happy&0.108 (0.105,0.111) \\ \hdashline
Emotional state&pms&0.086 (0.083,0.089) \\ \hdashline
Emotional state&sad&0.076 (0.073,0.079) \\
\hline
\end{tabular}
\end{center}
Kolmogorov-Smirnov test results for symptom tracking patterns that are significantly different (at a $p=0.000001$ level) between users in the consistently not highly variable and consistently highly variable groups.
\end{table*}

\paragraph*{Reported symptom differences---Women in the consistently highly variable group display more heterogeneous period tracking behavior.}
We find that women in the consistently highly variable group are significantly more likely not to report heavy periods throughout their cycle history (odds ratio of 1.734 on the low extreme end of the proportion range in Table~\ref{tab:interesting_low_extremes}). Additionally, the tracking pattern for spotting period flow is more heterogeneous for the consistently highly variable group, as shown by the higher odds ratios on both extremes of the proportion range, (\ie either in all or none of their cycle history) shown in Tables~\ref{tab:interesting_low_extremes} and~\ref{tab:interesting_high_extremes}.

\paragraph*{Reported symptom differences---Women in the consistently highly variable group report pain-related symptoms more unpredictably.}
We observe generally more heterogeneous experiences for non-bleeding related symptoms like pain for the consistently highly variable group. Of particular interest is the finding that users in the consistently highly variable group are much more likely associated with tracking headaches and tender breasts in at least 95\% of their cycles, with odds ratios of 1.663 and 1.715 respectively (see Table~\ref{tab:interesting_high_extremes}).

\begin{landscape}
\begin{table*}
\caption{Pain and period tracking odds ratios for low extreme end of the proportion range}
\label{tab:interesting_low_extremes}
\begin{center}
\begin{tabular}{|c|c|c|c|c|}
\cline{1-5}
 & & Consistently highly variable group's& Consistently not highly variable group's& Odds ratio (95\% CI) \\
Category & Symptom & likelihood for $\lambda_s < 0.05$ (95\% CI) & likelihood for $\lambda_s < 0.05$ (95\% CI) &for $\lambda_s < 0.05$ \\ \cline{1-5}
Period flow&heavy&0.170 (0.169,0.170) &0.098 (0.096,0.100) &1.734 (1.703,1.766) \\ \hdashline
Period flow&spotting&0.314 (0.313,0.315) & 0.239 (0.237,0.241) &1.314 (1.300,1.328) \\ \hdashline
Type of pain experienced&headache&0.326 (0.325,0.327) &0.269 (0.266,0.272) &1.212 (1.199,1.225)  \\ \hdashline
Type of pain experienced&tender breasts&0.366 (0.365,0.367) &0.320 (0.317,0.322) &1.145 (1.134,1.156) \\
\cline{1-5}
\end{tabular}
\end{center}
	Likelihood of low $\lambda_s$ per group, with the associated odds ratio (and 95\% confidence intervals). The probability of not tracking `heavy period' for users in the consistently highly variable group is 0.17 and 0.098 in the other, with an odds ratio of 1.734: the consistently highly variable group is more likely not to track `heavy period'.
\end{table*}
\begin{table*}
\caption{Pain and period tracking odds ratios for high extreme end of the proportion range}
\label{tab:interesting_high_extremes}
\begin{center}
\begin{tabular}{|c|c|c|c|c|}
	\cline{1-5}
	 & & Consistently highly variable group's& Consistently not highly variable group's & Odds ratio (95\% CI) \\
	Category & Symptom & likelihood for $\lambda_s> 0.95$ & likelihood for $\lambda_s> 0.95$ &  for $\lambda_s> 0.95$ \\ \cline{1-5}
Period flow&heavy& 0.078 (0.077,0.079) & 0.096 (0.094,0.097) &0.817 (0.802,0.833) \\ \hdashline
Period flow&spotting& 0.067 (0.066,0.067) &0.039 (0.037,0.040) &1.729 (1.679,1.782) \\ \hdashline
Type of pain experienced&tender breasts& 0.193 (0.192,0.194) &0.113 (0.111,0.115) & 1.715 (1.684,1.746) \\ \hdashline
Type of pain experienced&headache& 0.218 (0.217,0.219) &0.131 (0.129,0.133) &1.663 (1.636,1.691) \\
	\cline{1-5}
\end{tabular}
\end{center}
Likelihood of high $\lambda_s$ per group, with the associated odds ratio (and 95\% confidence intervals). The probability of consistently tracking `tender breast' pain for users in the consistently highly variable group is 0.193 and 0.113 in the other, with an odds ratio of 1.715: the consistently highly variable group is more likely to regularly track  `tender breast' pain.
\end{table*}
\end{landscape}
\section{Discussion}
\label{sec:discussion}

Characterization of menstrual patterns has been previously explored, though typically in relation to cycle and period lengths only. While common knowledge refers to a 28-day cycle as ``normal,'' this belief has been consistently disproved by clinical studies~\cite{j-Treloar1967,j-Chiazze1968}, as well as by recent analysis of high-level cycle characteristics via menstrual self-tracking apps~\cite{j-Bull2019,j-Symul2019}.

Overall, our results align with conclusions from these studies in that the cycle lengths have slightly higher values (median of 29 in our dataset) and wider ranges than what was previously commonly believed. While our study population demographics may differ slightly from other studies, we believe these still provide a reasonable basis for comparison. We show comparative summary statistics in
\ifx\includesi\undefined \href{https://github.com/iurteaga/menstrual_cycle_analysis/blob/master/doc/characterization/supplementary_information.pdf}{the Supplementary Information}\else the Supplementary Information\fi
, demonstrating the consistency of our cycle and period characteristics: ($i$) the average number of cycles tracked per user in this dataset (12.9) is bigger than in~\cite{j-Bull2019} (8.6), while it matches those (12.8) of~\cite{j-Symul2019}; ($ii$) the cycle length statistics are all similar: mean of $29.7$ in this work, $29.3$ in~\cite{j-Bull2019}; median of 29 in this work, 28 in~\cite{j-Symul2019}. Interestingly, this work and~\cite{j-Bull2019} both report an overall variability of cycle length of around 5 days, and this work and~\cite{j-Symul2019} both acknowledge the presence of ``a heavy tail on longer cycles.'' The period length averages of this work and~\cite{j-Bull2019} are in agreement as well ($4.08 \pm 1.76$ vs. $4.0 \pm 1.50$, respectively). Besides, these high-level cycle statistics align well with results of previous clinical studies~\cite{j-Treloar1967,j-Chiazze1968,j-Creinin2004}.

Examining cycle length is often insufficient for capturing all fluctuations in menstrual patterns---studies regarding menstrual variability showcase that although average cycle length is associated with cycle length consistency, women still experience significant variability in cycle lengths regardless of their average cycle length~\cite{j-Creinin2004}. In this work, we address this limitation by utilizing our proposed metric, median CLD, to characterize menstrual cycle variability.
Separating users according to their median CLD yields two distinct groups of users with statistically significant differences in cycle length, cycle length variations, and symptom tracking behaviors. We are unaware of any single figure of merit which so helpfully separates users into distinct segments. Clue uses the International Federation of Gynecology and Obstetrics (FIGO) definitions for clinically irregular cycles in the app~\cite{Druet2018}, but has not found connections with differences in tracking.

While there has been ample work on hormone-level characterizations of the menstrual cycle~\cite{j-Johannisson1987, j-Fehring2006,j-Lenton1984,j-Lenton1984a}, studies of the relationship between menstrual patterns and symptomatic variables are limited---recent work has explored this association using self-tracked data, but over a limited set of symptoms~\cite{ip-Pierson2018} and without discriminating over age or birth control usage~\cite{j-Pierson2019}. A method for estimating ovulation timing based on Fertility Awareness Method observations (i.e., basal body temperature (BBT), cervical mucus, cervix position, and vaginal sensation) has been presented~\cite{j-Symul2019}, but such data is inaccessible for this study due to the European Union's General Data Protection Regulation and other data-privacy concerns (sensitive fields such as appointments, ovulation and pregnancy tests, and BBT were not available in Clue's dataset). Nonetheless, such studies showcase the potential large-scale self-tracked data offer in exploring questions relating to menstruation.

This work further demonstrates that mobile self-tracked data provide an accessible option for clinicians and researchers investigating changes of a variable of interest across the menstrual cycle. Our dataset allows us to explore symptoms of interest like pain, types of bleeding, and emotions explicitly, and we are able to connect variability in cycle lengths to patterns in self-reported symptom tracking. In contrast to existing work, our methods allow us to comment quantitatively and qualitatively on the menstrual experience over a broad set of symptoms. While cycle length has been proposed as a biomarker of menstrual health (e.g., very long and very short cycles are associated with a higher risk of infertility), this work suggests that cycle variability may also be a useful biomarker.
 
We propose a definition of menstrual cycle variability and find that users in high and low variability groups showcase both statistically significant differences in their cycle statistics as well as in their symptom tracking patterns. We argue that the discovery of such distinct forms of symptom expression allows for phenotype identification and the investigation of clinical associations. In particular, of the symptoms which show statistically significant association with timing data (as measured by median CLD), some of the most arguably unambiguous ones like period flow and pain are also the diagnostic symptoms frequently appearing in the assessment of menstrual health conditions like endometriosis and polycystic ovary syndrome (PCOS). Thus, cycle variability and these high-signal self-tracked symptom patterns can be potentially useful either for predicting each other (e.g., predicting cycle variability from symptoms) or health consequences (e.g., PCOS), insights which are useful to both clinicians and users.

Our perspective on how users experience their menstruation enables development of data-driven models to predict multiple aspects of the menstrual cycle based on self-tracked history, ranging from modeling time to next cycle, to forecasting occurrence of specific symptoms a user might report, to detecting underlying medical conditions. Equipped with the results of this work at the cycle level, future work will consist of identifying further differences at a finer grain, namely across the different menstrual phases. 

Despite the strength of these results, there are several mitigating factors to bear in mind. We acknowledge that self-tracked data may be unreliable for several reasons, such as inconsistent user engagement or ambiguous symptomatic language. For example, there is potential overlap between similar-sounding symptoms, \eg some users may track `low energy,' whereas others may track `exhausted.' Users can also engage inconsistently by tracking an unequal number of cycles or forgetting to track their period. We successfully ameliorate the latter issue by excluding unexpectedly long cycles utilizing our proposed procedure. For the former issue, we observe that the consistently highly variable group tracked a lower number of cycles on average (see Table~\ref{tab:statistics_cycles}). However, we note that the number of users who only tracked two cycles (after our preprocessing steps) is small across the entire cohort, representing 2.62\% and 0.57\% of the consistently highly and not highly variable groups, respectively.

In addition to the risk of inconsistent user engagement, inherent in the nature of self-tracked data is the challenge of disentangling user behavior from true physiological experiences. The design and selection of symptoms for the Clue app was based both on the scientific literature around the menstrual experience and research on which categories users deemed important. As such, in order to encompass a wide range of relevant menstrual and health experiences, the available tracking categories are broad and treated with equal importance. However, we acknowledge that since the symptoms in the app are not based on validated scales and are not designed for diagnosis of specific conditions, they are most likely not granular (nor targeted) enough to make definite claims about specific conditions. Furthermore, while there are infotexts in the Clue app that explain each tracking category, self-reported data is influenced by individual user interpretation and by how users use the app to meet their own needs; we cannot guarantee that each category has the same meaning for each user.

In this paper, we take symptom tracking behavior to be a proxy for true physiological behavior. However, we are cognizant of the fact that these are not necessarily equivalent. Note that it is very difficult to know what the true physiological experience is in any circumstance: e.g., the experience of menopause varies greatly by culture~\cite{j-Jones2012}. With self-tracked data and without access to ground truth, it is complicated (if not impossible) to truthfully distinguish the experienced symptoms from the tracked ones, due to the presence of engagement artifacts and other unforeseen factors. As such, we have taken steps to reduce tracking artifacts with preprocessing techniques, but recognize that limitations remain. Nonetheless, it remains useful to examine these datasets to better understand not only women's menstrual experiences at scale, but also how to improve self-tracking technologies to enable clearer, more interpretable datasets in the future.

Overall, large-scale self-tracked mobile-health data allow us to quantitatively explore the question of characterizing menstrual behavior. Our findings reinforce the claim that menstruation is characterized by variability rather than by regularity~\cite{j-Arey1939,j-Treloar1967,j-Chiazze1968,j-Creinin2004,j-Vitzthum2009}.
We find variation in cycle length statistics as well as in self-reported symptoms, showcasing the spectrum of how women experience their menstruation. We reveal statistically significant relationships between the variability of cycle length and self-reported qualitative symptoms.
The identified set of symptoms which show association with timing data (e.g., period flow and pain) are the diagnostic symptoms frequently leveraged for diagnosis of health-relevant conditions, such as endometriosis and PCOS, insights that are useful to both clinicians and users.
More broadly, we also develop a methodology for identifying artifacts in self-tracked data, which can be extended to other self-reported menstrual tracking datasets. This work not only statistically verifies the variation of menstrual experience, but also presents promising opportunities for future statistical modeling, prediction, and the potential to inform diagnosis of menstrual-related disorders.

\clearpage 
\section{Methods}
\label{sec:methods}

\subsection*{Data overview}
\label{ssec:methods_data}
We leverage a de-identified version of the Clue data warehouse, a dataset of $117,014,597$ self-tracking events for $378,694$ users in our cohort of interest.
Clue app users input overall personal information at sign up, such as age and hormonal birth control (HBC) type. The dataset contains information from 2015-2018 for users worldwide, covering countries within North and South America, Europe, Asia and Africa (see 
\ifx\includesi\undefined \href{https://github.com/iurteaga/menstrual_cycle_analysis/blob/master/doc/characterization/supplementary_information.pdf}{the Supplementary Information }\else the Supplementary Information \fi
for a detailed count of cohort users per country). Users can self-track symptoms over time across the 20 available categories (see Table~\ref{tab:statistics_symptoms} for symptom list) and can pre-select which categories they wish to track when they sign up---all users do not track all categories. 

Clue app users track an event by selecting a category and then choosing an associated symptom. Each row in the primitive dataset represents a tracked event $e$, with relevant fields being ($i$) the user $u$ that tracked the event $e_u$, ($ii$) the reported symptom $s$ in that event $e_u=s$, and ($iii$) the user-specific cycle $c_e$ in which the event takes place. A menstrual cycle is defined as the span of days from the first day of a period through to and including the day before the first day of the next period ~\cite{j-Vitzthum2009}. A period consists of sequential days of bleeding (greater than spotting and within ten days after the first greater than spotting bleeding event) unbroken by no more than one day on which only spotting or no bleeding occurred. Note that Clue considers a menses duration longer than 10 days as an outlier, as it would exceed mean period length plus 3 standard deviations for any studied population~\cite{j-Vitzthum2009}. In addition, a user has the opportunity to specify whether a cycle should be excluded from their Clue history---for instance, if the user feels that the cycle is not representative of their typical menstrual behavior due to a medical procedure or changes in birth control.
\subsection*{Cohort definition}
\label{ssec:methods_cohort}
A cohort of users and cycles was selected for this analysis, based on factors including age, HBC usage, cycle length, and engagement patterns. Recall that we restricted our data to users aged between 21-33 years, since menstrual cycles are relatively less variable in length and more likely to be ovulatory during this age range~\cite{j-Treloar1967,j-Chiazze1968,j-Ferrell2005,j-Vitzthum2009,j-Harlow2012}.
At younger ages, the reproductive axis (the hypothalamic-pituitary-ovarian axis) in some women, especially those who experienced a later than average age at menarche, may not be fully matured. At older ages, some women may be experiencing premature menopause. Restricting our sample to this age group substantially reduces the influence of confounders like undetected heterogeneity on our analyses. Per-age details like cycle and period length statistics are provided in 
\ifx\includesi\undefined \href{https://github.com/iurteaga/menstrual_cycle_analysis/blob/master/doc/characterization/supplementary_information.pdf}{the Supplementary Information}\else the Supplementary Information\fi.

\begin{figure}[!t]
	\centering
	\includegraphics[width=0.75\columnwidth]{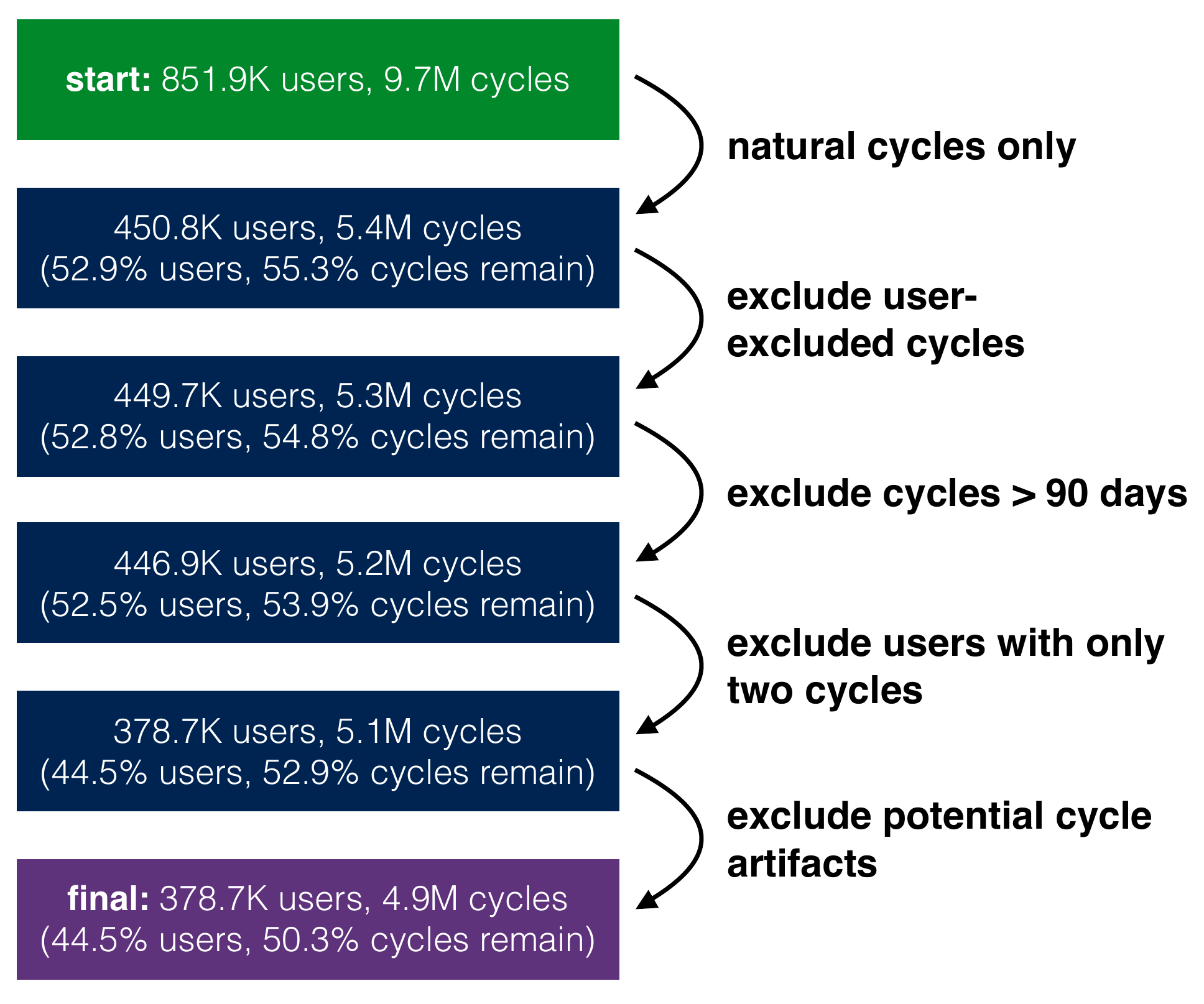}
	\caption{Step-by-step filtering process for computing the final user and cycle cohort. The percentage of users and cycles removed at each step is computed out of the initial numbers. Note that we only include users aged between 21-33 years, since women exhibit more stable menstrual behavior in their `middle life' phase~\cite{j-Treloar1967,j-Chiazze1968,j-Ferrell2005,j-Vitzthum2009,j-Harlow2012}.}
	\label{fig:data cohort diagram}
	\vspace*{-2ex}
\end{figure}

Since HBC and copper IUDs have been shown to impact cycle length and other aspects of menstruation, we consider natural menstrual cycles only. Therefore, we ignore cycles from users who reported some form of HBC (patch, pill, injection, ring, implant) or IUD (we excluded all cycles with evidence of IUD use, as there is no explicit distinction between hormonal and copper IUD usage in the dataset). This step removes about 45\% of the cycle data, but is crucial to studying menstruation in a standardized way across users, else it would be unclear whether an exhibited menstrual behavior was due to physiology or the effect of birth control. We exclude cycles that a user deems to be anomalous to avoid potential artifacts in cycle patterns. In addition, we eliminate cycles greater than 90 days long, as well as users who have only tracked two cycles, to rule out cases that we argue indicate lack of engagement or non-continuous use of the app. Finally, we exclude cycles where we believe the user forgot to track their period, hence resulting in an artificially long cycle length; we explain this procedure below. The effect of these filtering steps on the dataset is outlined in Figure~\ref{fig:data cohort diagram}, with the final step indicating the removal of aforementioned suspected artificially long cycles. In total, the proposed data filtering steps reduced the size of the cycle dataset by about 49\%. However, the resulting age-specific, natural cycle-only user cohort and corresponding dataset with potential artifacts removed enables us to study our research questions in a less noisy setting.

\subsection*{Ethics}
\label{ssec:ethics}

The research presented here was exempt from Columbia University IRB approval, in accordance with 45CFR46.101(b), as all data is de-identified and no participant risks are associated with taking part in the study. Participants do not receive direct benefit from this study, but their participation contributes to the general knowledge of menstrual cycles and their symptoms.
\subsection*{Characterizing longitudinal menstrual tracking via cycle length difference}
\label{ssec:methods_cld}

There are many useful ways to characterize menstrual cycles, each of which offers its own advantages and disadvantages. For instance, cycle length provides insight into the length of time between periods and has been widely documented to vary across women~\cite{j-Chiazze1968,j-Muenster1992,j-Belsey1997,j-Burkhart1999,j-Vitzthum2000,j-Creinin2004,j-Williams2006}, but is insufficient for understanding menstrual cycle length volatility, as it fails to characterize variability from one cycle to the next.

\begin{figure}[!h]
	\centering
	\includegraphics[width=0.75\columnwidth]{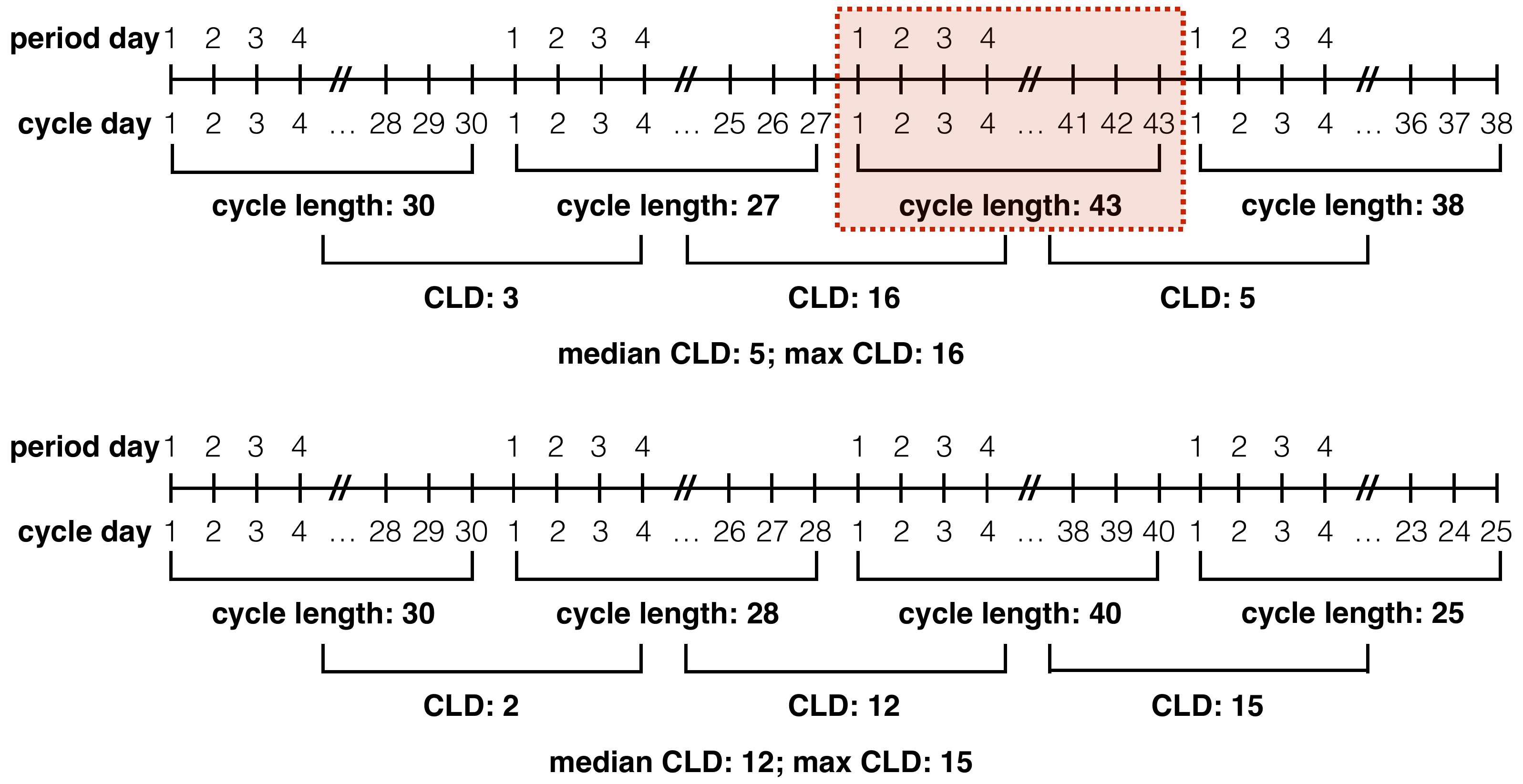}
	\caption{We provide illustrative examples of identifying a cycle tracking artifact (top) and characterizing a user's regularity (bottom) based on CLD statistics. In each example, we display a user's cycle history with a total of four cycles. Cycle length is computed as the length of time between the first day of a period and the first day of the next period, and CLD is computed as the absolute difference between subsequent cycle lengths (i.e., if a user has $n$ cycles tracked, they will have $n-1$ CLD values). Period length is computed by counting the number of sequential days on which there is menstrual bleeding greater than spotting (`light,' `medium,' or `heavy'). Two such sequences are considered one period if separated by no more than one day of non-bleeding/spotting. In the top example, the user's second CLD exceeds their median by at least 10, and thus we identify the corresponding `artifically long' cycle in red---this cycle will be excluded from our analysis. In the bottom example, the user's median CLD is at least 9, and thus it will be classified as a consistently highly variable user.}
	\label{fig:data tracking}
	\vspace*{-2ex}
\end{figure}

We propose computing cycle length differences (CLDs), which we define as the absolute differences in subsequent cycle lengths. CLDs represent a user's longitudinal cycle tracking history by quantifying their between-cycle volatility. This metric captures menstrual patterns regardless of specific cycle lengths, allowing us to measure fluctuation over time and identify those who are consistently highly variable. This metric does not capture some other menstrual phenomena, such as cycle lengths growing at a constant pace---that is, if a cycle length grew consistently by two days with each cycle, the CLDs would all be equal to two, but there would be a large difference between the shortest and longest cycle lengths. However, CLDs and related metrics of median and maximum CLD do allow us to characterize users on the extreme ends of the between-cycle variability spectrum and identify potential cycle tracking artifacts, as described in the following sections. Figure~\ref{fig:data tracking} outlines the computation of CLDs and related statistics.

\subsection*{Quantifying engagement with cycle tracking}
\label{ssec:methods_engagement}
We propose a methodology for identifying cycles associated with lack of app engagement, specifically where users forgot to track their period, since this may inflate the corresponding computed cycle length. Our procedure allows us to distinguish physiological behavior (\ie true `long' cycle lengths) from tracking artifacts (\ie artificially inflated cycle lengths), which allows us to more reliably utilize symptom tracking behavior as a proxy for true physiological behavior.

\begin{figure}[!h]
	\centering
	\includegraphics[width=0.75\columnwidth]{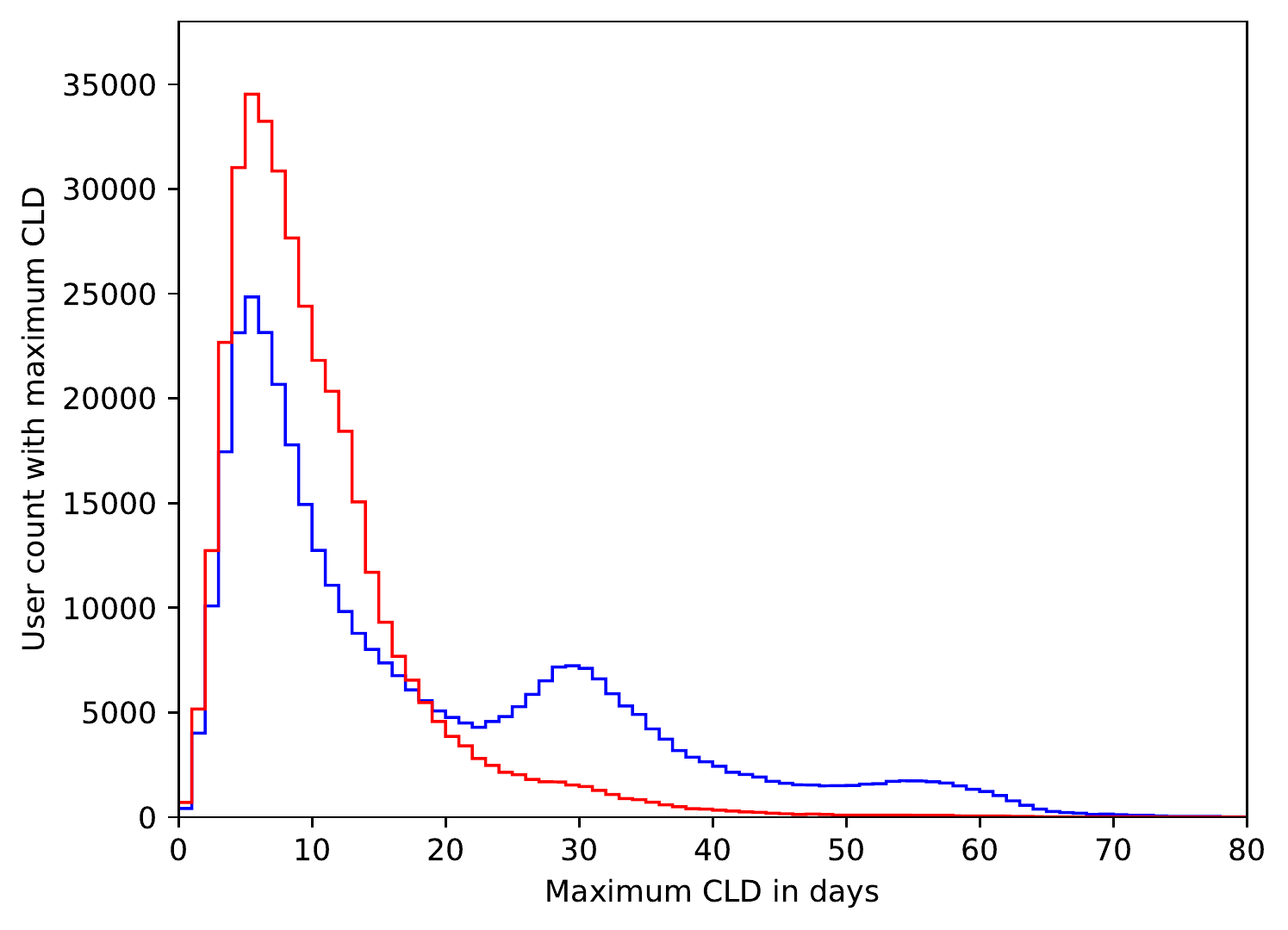}
	\caption{For each user, we compute the maximum CLD and plot a histogram before (blue) and after (red) excluding cycles without user engagement (i.e., cycles that are potential artifacts). We see that the multi-modal behavior (peaks at around 30 and 60 days) is largely dampened upon removing these cycles. In addition, the fat right-hand tail in the red curve implies that we preserve the natural variation in cycle length---we are not simply removing long cycles.}
	\label{fig:hist max intercycle length}
	\vspace*{-2ex}
\end{figure}

\begin{figure}[!h]
	\centering
	\includegraphics[width=0.75\columnwidth]{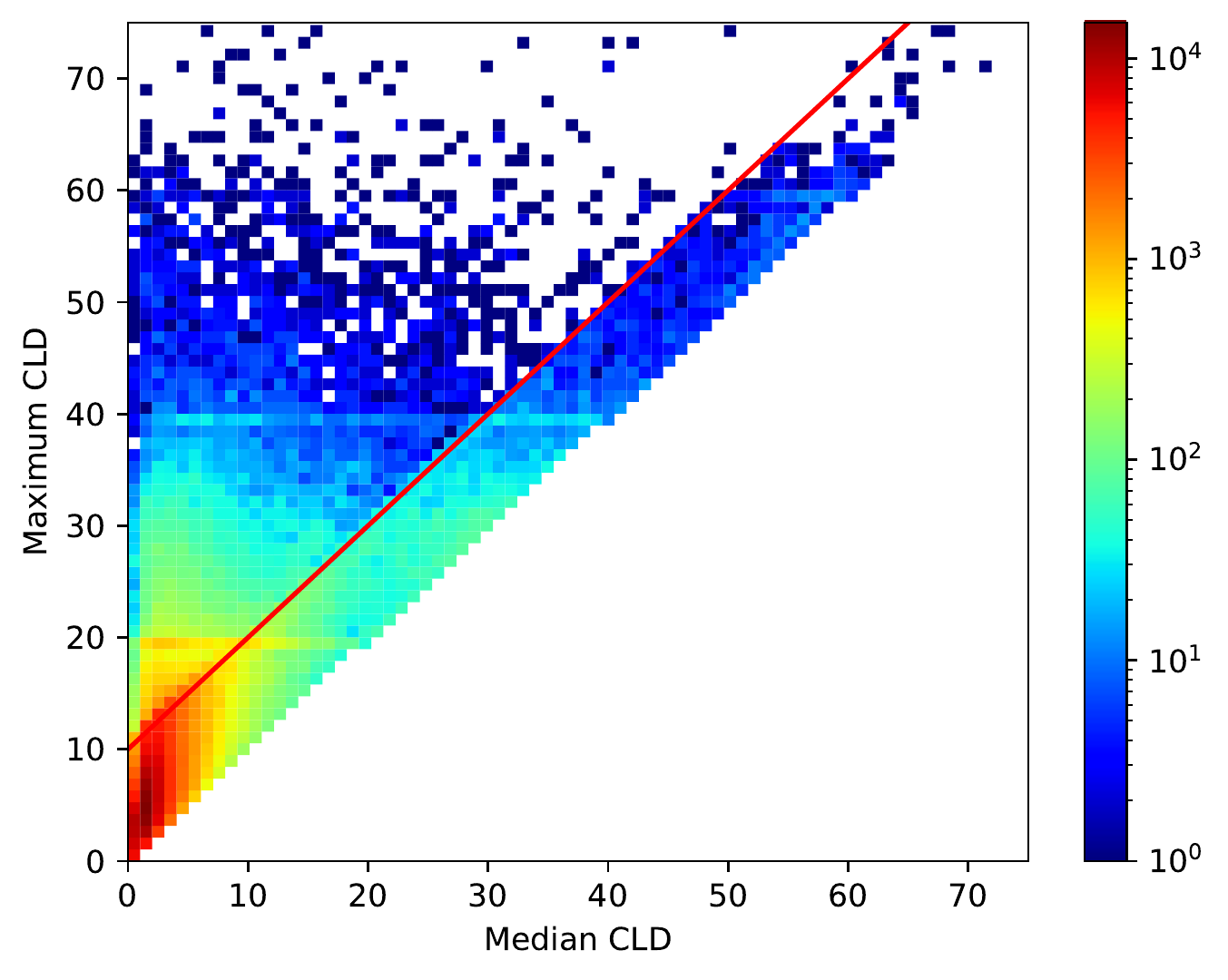}
	\caption{We plot a two-dimensional histogram of users' median CLD versus maximum CLD in logarithmic space, as well as the line where maximum CLD is equal to median CLD plus 10 in red. We can see that the line separates out a highly concentrated region of users, as well as a more scattered region of users. Specifically, the majority of the mass falls under this line, as showcased by the concentrated red color in the lower lefthand corner of the plot and a diagonal band extending upwards, while the region above the line is more spread out. Thus, we examine the cycles that fall above the line as possible cycle tracking artifacts.}
	\label{fig:median vs max}
	\vspace*{-2ex}
\end{figure}

Figure~\ref{fig:hist max intercycle length} showcases how maximum CLD impacts our overall picture of user engagement. In particular, the multi-modal nature of the histogram of maximum CLD (in blue) indicates that there may be cycles where users forgot to track their period, resulting in an overestimation of cycle length. Note the peaks around 30 and 60 days, which may correspond to users forgetting to track one or two periods, respectively. That is, consider a user who exhibits a perfectly uniform cycle length of 30 and hence always has a CLD of 0. If this user were to forget to track a period once in their history, then the app would record that they have a cycle length of 60 and a maximum CLD of 30---such a user would fall into the first peak of the histogram. The discrepancy between regular patterns (via the median CLD) and extreme events (maximum CLD) is further illustrated in Figure~\ref{fig:median vs max}.


We identify cycles that are `atypically long' compared to the `typical' cycle length for each user by examining the difference between each CLD and the median CLD of that user. An illustrative example is provided in the top panel of Figure~\ref{fig:data tracking}, where the third cycle appears to be `atypically long.'
Specifically, we flag cycles per user where the corresponding CLD exceeds the user's median CLD by at least 10 days as `atypically long' (the longer of the two cycles corresponding to the CLD is flagged). This cutoff is based on an attempt to find in the data, rather than posit a priori, a feature that would distinguish `typical' from `extreme' (i.e., abnormally long) reported cycles. To do so, we plot the two-dimensional histogram (see Figure~\ref{fig:median vs max}) of maximum CLD against median CLD; this is a histogram in which every example is a user. We observe a clear visual feature: a band of users we consider `typical,' for whom maximum CLD was within 10 days of the median CLD, and a scatter of other users for whom their maximum CLD could be far larger. To capture this visually striking feature (see the diagonal red line along maximum CLD equal to 10 more than median CLD in Figure~\ref{fig:median vs max}), we defined extreme events as those at least 10 days above the median CLD. We consider the cycles flagged as `atypically long' to be the result of cycle engagement artifacts and exclude them from our analysis.

As shown in Figure~\ref{fig:hist max intercycle length} (red line), the multi-modal shape is largely removed after eliminating the `atypically long' cycles. We find that 42\% of the cohort has at least one such cycle, and for these users we exclude a small number (1.59) of cycles per user on average. This indicates that our method is stringent enough to identify artificially long cycles, but conservative enough to preserve the heterogeneity of the data.

We further validate our method by examining tracking activity during the interval where a user is expected to track their period for each of these excluded cycles and find that in 89.18\% of such cases there is no evidence of bleeding-related events during this interval, i.e., the user likely did not engage in period tracking. We define this interval as the user's last reported period day plus their median cycle length, plus or minus their median period length. In the remaining 10.82\% of excluded cycles, it is unclear whether the bleeding-related events tracked during this interval represent a period or some other non-period bleeding. Note that by our definition of a period, a single bleeding event is not synonymous with period. As a conservative measure and to maintain consistency of our definitions for period and artificially long cycles, we exclude those cycles from our analysis. This ensures a coherent data pre-processing pipeline and impacts the results minimally (these excluded cycles with some bleeding-related events amount to only 0.56\% of all cycles). Quantifying inconsistent tracking engagement allows us to ameliorate its impact on subsequent analyses.
\subsection*{Characterizing users according to cycle length variability}
\label{ssec:methods_variability}
\begin{figure}[!h]
\centering
\includegraphics[width=0.75\linewidth]{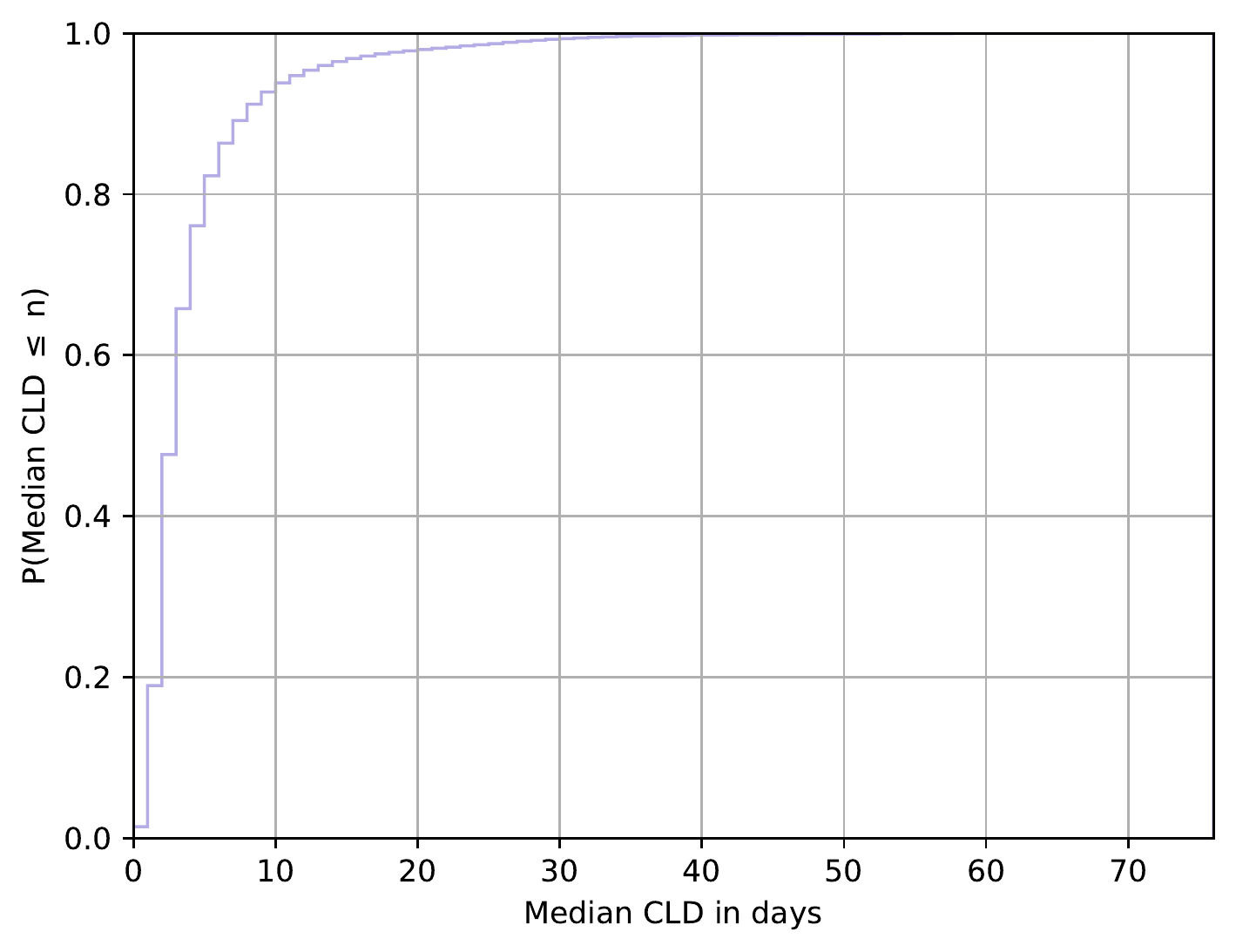}
\caption{Looking at the cumulative distribution of median CLD, we see that the curve flattens out significantly around the `elbow' at 9 days; thus, we choose greater than 9 days as our cutoff for our definition of consistently highly variable.}
\label{fig:median intercycle length}
\vspace*{-2ex}
\end{figure}

We acknowledge that there is a wide spectrum of variability in women's menstrual health experiences, and we wish to examine those who fall at opposite ends of the variability spectrum on the basis of their cycle pattern consistency. We choose the median CLD metric for our analysis, as it is robust to outliers (the mean would be more susceptible to being skewed by rare events). Upon examining the cumulative distribution of this metric across users in Figure \ref{fig:median intercycle length}, we consider a median CLD of greater than 9 days to be an appropriate stringent cutoff for identifying consistently highly variable menstrual patterns. This choice aligns with previous work on menstrual pattern analysis: cycle length variability studies in Guatemalan, Bolivian, Indian, U.S. and European women noted differences in the maximum and minimum cycle length ranging from 6 to 14 days~\cite{j-Muenster1992,j-Belsey1997,j-Burkhart1999,j-Vitzthum2000,j-Creinin2004,j-Williams2006}.
Our proposed cutoff separates users into two distinct groups of menstrual patterns: the vast majority (92.32\%) of the population falls to the left of this threshold, and thus the consistently highly variable group (the remaining 7.68\%) represent those whose variability is extreme. As discussed in the Results section, we observe that the highly variable group experiences more drastic fluctuations in cycle length. We confirm that the cycle length distributions differ significantly between the two groups using a two-sample Kolmogorov-Smirnov test~\cite{j-Kolmogorov1933}.

\subsection*{Quantifying symptom tracking behavior across user groups}

We focus on symptom tracking behavior at the cycle level, evaluating how often throughout their longitudinal tracking users track each symptom, regardless of when within the cycle the tracking occurred (\ie ignoring at which phase or day of the cycle the symptom occurred). Note that because cycle length varies both within each user's longitudinal tracking and across women, the number of tracking events per cycle would be skewed by cycle length.
To combat this issue, we measure the per-user proportion of cycles where a symptom has been tracked.  

We also want our metric to capture symptom tracking behavior for cycles where users were interested in tracking the associated category (recall how users do not have to track all categories). Specifically, our analysis focuses on how often a user $u$ has a symptom $s$ tracking event $e_u=s$ per cycle $n$, given that they have tracked symptoms within the associated category $C$ at least once across all their cycles $N_u$. We refer to this metric as the \textit{`proportion of cycles with symptom out of cycles with category,'} mathematically denoted as
\begin{equation}
\lambda_{us} = \frac{\sum_{n=1}^{N_u}\mathds{1}[\exists e_u=s]}{\sum_{n=1}^{N_u}\mathds{1}[\exists e_u \in C]} \; .
\label{eq:our_metric}
\end{equation}
That is, to account for whether a user is actually interested in the symptom at hand, we compute the proportion of cycles with a symptom being tracked out of the number of cycles where the user has tracked the category related to that symptom. For example, consider a user who tracked 8 cycles; out of these, she tracked any of the symptoms within the pain category for 5 cycles. For 1 of these cycles only, she tracked the symptom `headache,' while for 4 of these cycles, she tracked `tender breasts.'
Our metric $\lambda_s$ captures the tracking regularity of a given symptom across a user's cycles. When applied to the example user, 20\% of cycles with pain have `headache' tracked, while 80\% of the same cycles have reports of `tender breasts.'
In essence, $\lambda_{us}$ is the conditional probability that user $u$ tracks the specific symptom $s$ given that she has tracked any symptom from the symptom's corresponding category. Our metric measures per-cycle symptom tracking frequency and is robust to ($i$) different cycle lengths and number of cycles (as it is normalized with respect to each user’s number of cycles), ($ii$) different user app interests (as it is contingent on whether the user has shown interest in tracking such category at least once), and ($iii$) different app usage behaviors (as it does not depend of how many times within a cycle the symptom is tracked).

We study the cumulative distributions of $\lambda_{s}$ per group (\ie $\lambda_{us}$ for all users $u$ within each variability group), as well as how such densities are different on their support boundaries across groups. Since we lack a mechanistic model of what distribution the data might follow and wish to use a test meaningful for any distribution, we utilize a nonparametric test suitable for any ordinal (as opposed to, e.g., binary or categorical data): the Kolmogorov--Smirnov (KS) test~\cite{j-Kolmogorov1933}. This test of the comparative equality of one-dimensional probability distributions arising from two samples allows us to quantify statistical differences in symptom tracking behavior between groups. Specifically, we compare the distributions of the proportion of cycles with a symptom tracked (out of cycles with its corresponding category), for the consistently not highly variable and consistently highly variable user groups. The KS statistic quantifies the distance between the empirical cumulative distributions of two samples, and the associated test is sensitive to differences in both location and shape of said distributions, allowing us to characterize \textit{where} and \textit{how much} the symptom tracking patterns (as measured by the proposed $\lambda_s$ metric) differ between groups. 

In the two-sample case, the null distribution of the KS statistic is calculated under the null hypothesis that the samples are drawn from the same distribution, where the distribution considered under the null hypothesis is an unrestricted continuous distribution (i.e., no distributional assumption is made on the symptom tracking patterns). The KS statistic depends on the number of data points within each of the populations (i.e., the number of observations that we have for each group when computing their per-symptom empirical cumulative density function). The null hypothesis is rejected at level $\alpha$ if
\begin{equation}
D_{n,m}>\sqrt{-{\frac {1}{2}}\ln \alpha \cdot \frac {n+m}{nm}} \; ,
\end{equation}
where $n$ and $m$ are the sizes of the first and second data samples respectively, and $D_{n,m}$, the computed two-sample KS statistic. The $p$-values reported by the KS test consider observed sample sizes, accounting for the impact of whether certain symptoms are more or less frequently logged in each group.

In order to explore \textit{how} the empirical distributions differ, we study their support boundaries, \ie $p(\lambda_s> 0.95)$ and $p(\lambda_s < 0.05)$. These represent how likely users in each group are to either consistently track a symptom throughout their cycle history (\ie in almost every cycle where they track the category), or to not track it at all (\ie in very few of their cycles where they track the category). We compute the odds ratio of these values (on either the high extreme or low extreme end of the proportion range) for the consistently highly variable group to the consistently not highly variable group. If we have an odds ratio greater than 1 for the high extreme end of the metric range for a symptom, this would indicate that the consistently highly variable group is more likely to report a very high proportion of cycles with that symptom. On the other hand, an odds ratio greater than 1 for the low extreme end of the proportion range (\ie the proportion of cycles with a symptom tracked is close to zero) indicates that the consistently highly variable group is more likely \textit{not} to report such a symptom.

When possible, 95\% confidence intervals have been added to reported KS values using bootstrap analysis. To do so, we draw 100,000 random samples---resampled with replacement---from each variability group and report the estimated mean KS statistic values and their 2.5 and 97.5 percentiles. 

\section{Data availability}
\label{sec:data availability}
The database that supports the findings of this study was made available by Clue by BioWink. While it is de-identified, it cannot be made directly available to the reader. Researchers interested in gaining access to the data can contact Clue by BioWink and establish a data use agreement with them.  
\section{Code availability}
\label{sec:code availability}
Our code has been developed using open source tools in Python with common statistical libraries (e.g., Pandas and SciPy). The code required for data pre-processing and producing results is available in the public GitHub repository \href{https://github.com/iurteaga/menstrual_cycle_analysis}{https://github.com/iurteaga/menstrual\_cycle\_analysis}.
\section{Acknowledgements}
The authors are deeply grateful to all Clue users whose de-identified data have been used for this study.

\section{Author contributions}
KL and IU contributed equally to this work. KL, IU, CHW, and NE conceived the proposed research and designed the experiments. KL and IU processed the dataset, conducted the experiments, and wrote the first draft of the manuscript. CHW, AD, AS, VJV, and NE reviewed and edited it. All authors read and approved the manuscript.
\section{Competing interests}
KL is supported by NSF's Graduate Research Fellowship Program Award \#1644869. IU, CHW and NE are supported by NSF Award \#1344668.
KL, IU, CW, and NE declare that they have no competing interests. AD, AS and VJV were employed by Clue by BioWink at the time of this research project.

\clearpage

\bibliographystyle{abbrvnat}

\ifx\includesi\undefined
\else
\clearpage
\input{supplementary/si}
\fi
\end{document}